\documentclass[12pt]{article}
\usepackage{ragged2e}
\usepackage{jheppub}
\usepackage{dsfont}
\usepackage{amsmath,amssymb,amscd,amsfonts,mathtools}
\usepackage{xcolor}
\definecolor{markgreen}{RGB}{230,243,230}
\definecolor{darkolivegreen}{rgb}{0.33, 0.42, 0.18}
\definecolor{darkpastelgreen}{rgb}{0.01, 0.75, 0.24}
\DeclareMathOperator{\Tr}{Tr}

\DeclareMathOperator{\arccosh}{arccosh}
\DeclareMathOperator{\arcsinh}{arcsinh}
\DeclareMathOperator{\arctanh}{arctanh}

\usepackage[toc,page]{appendix}
\usepackage{epsfig}
\usepackage{epstopdf}
\usepackage{latexsym}
\usepackage{graphicx}
\usepackage{booktabs}
\usepackage{bbm}
\usepackage{color}
\usepackage{physics}
\usepackage{tensor}
\usepackage{verbatim}
\usepackage{subfig}
\usepackage{tikz}
\usepackage{ifthen}
\usetikzlibrary{matrix}
\usetikzlibrary{decorations.markings,calc,shapes,decorations.pathmorphing}
\usetikzlibrary{patterns}
\usetikzlibrary{positioning}

\pdfoutput=1

\newboolean{showcomments}
\setboolean{showcomments}{false}
\newcommand\rem[1]{\ifthenelse{\boolean{showcomments}}{{#1}}{}}
\newcommand{\be}{\begin{equation}}
\newcommand{\ee}{\end{equation}}

\title{\Large Holographic BCFTs and Communicating Black Holes}
\author{Hao Geng$^a$, Severin L\"ust$^b$, Rashmish K. Mishra$^b$ and David Wakeham$^c$}
\affiliation{$^a$Department of Physics, University of Washington, Seattle, WA, 98195-1560, USA.}
\affiliation{$^b$Harvard University, 17 Oxford St., Cambridge, MA, 02138, USA.}
\affiliation{$^c$Department of Physics and Astronomy, University of British Columbia,
6224 Agricultural Road, Vancouver, B.C. V6T 1Z1, Canada.
}

\emailAdd{hg666@uw.edu}
\emailAdd{sluest@g.harvard.edu}
\emailAdd{rashmishmishra@fas.harvard.edu}
\emailAdd{daw@phas.ubc.ca}

\abstract{We study the AdS/BCFT
duality between two-dimensional conformal field theories with two boundaries and three-dimensional anti-de Sitter space with two Karch-Randall branes.
We compute the entanglement entropy of a bipartition of the BCFT,
on both the gravity side and the field theory side.
At finite temperature this entanglement entropy characterizes the communication between two braneworld black holes, coupled to each other through a common bath.
We find a Page curve consistent with unitarity.
The gravitational result, computed using double-holographically realized quantum extremal surfaces, matches the conformal field theory calculation.
\\
\hspace{15pt}At zero temperature, we obtain an interesting extension of the AdS$_3$/BCFT$_2$ correspondence.
For a central charge $c$, we find a gap $(\frac{c}{16},\frac{c}{12})$ in the spectrum of the
scaling dimension
$\Delta_{\text{bcc}}$ of the boundary condition changing operator (which interpolates
mismatched boundary conditions on the two boundaries of the BCFT).
Depending on the value of  $\Delta_{\text{bcc}}$, the gravitational dual is either a defect global AdS$_3$ geometry or a single sided black hole, and in both cases there are two Karch-Randall branes.
}

\begin{document}	
\maketitle
\flushbottom

\section{Introduction}
Recent approaches to understanding
the black hole information problem using the AdS/CFT correspondence \cite{Penington:2019npb,Almheiri:2019psf} 
emphasize the role of \textit{entanglement islands} in restoring the unitarity of the underlying dynamics. 
Models of entanglement islands in higher dimensions are most cleanly realized in the Karch-Randall (KR) braneworld \cite{Karch:2000ct,Karch:2000gx}, later refined to the so-called AdS/BCFT correspondence \cite{Takayanagi:2011zk,Fujita:2011fp}. In these models \cite{Geng:2020qvw,Almheiri:2019psy} (see also \cite{Krishnan:2020fer,Ling:2020laa,Chen:2020hmv,Chen:2020uac,Chen:2020tes,Bousso:2020kmy,Hernandez:2020nem,Bhattacharya:2020uun,Chandrasekaran:2020qtn,Chou:2020ffc,Almheiri:2019hni,Rozali:2019day,Bak:2020enw,Cooper:2019rwk,Laddha:2020kvp,Chowdhury:2020hse,Basak:2020aaa,Kawabata:2021hac,Wang:2021woy,Deng:2020ent,Caceres:2020jcn,May:2021zyu,Karlsson:2021vlh,Karananas:2020fwx,Goto:2020wnk,Matsuo:2020ypv,Verheijden:2021yrb,Anderson:2021vof,Bhattacharya:2021jrn,Wang:2021mqq,Aalsma:2021bit,Kim:2021gzd,Hollowood:2021nlo,Ghosh:2021axl,Neuenfeld:2021bsb,Kawabata:2021vyo,Reyes:2021npy,Akal:2020twv,Akal:2021foz} for related studies), the black hole is living on the KR brane and is coupled to a finite temperature bath modeled by a conformal field theory (CFT) on a manifold with a boundary. The bath absorbs the radiation from the black hole and the entanglement island of a subsystem of the bath is the component of its entanglement wedge disconnected from it. The entanglement island lives on the KR brane and contains the black hole interior.

In these early models, the bath was non-gravitating, but a recent generalization of the KR construction involves gravitating baths \cite{Geng:2020fxl}. Several new features of entanglement islands emerge in this setup, and in particular, it becomes possible 
to study the information transfer between two black holes using the recently proposed \textit{wedge holography} \cite{Bachas:2017rch,Akal:2020wfl,Miao:2020oey}.  
Here, ``information transfer" is characterized by the entanglement entropy between complementary subsystems of the wedge-holographic field theory dual.

In \cite{Sully:2020pza,Rozali:2019day}, entanglement islands were studied for a three-dimensional bulk geometry, with a non-gravitating bath. Interestingly, in these cases the entanglement entropy can be computed using \textit{both} the bulk gravitational description and its dual boundary conformal field theory (BCFT) description. On the gravity side, the entanglement entropy is computed using the \textit{Ryu-Takayanagi} (RT) surface \cite{Ryu:2006bv,Ryu:2006ef}, while on the CFT side, using correlation functions of twist operators \cite{Calabrese:2009qy}. These two computations match each other under a natural extension of the conditions for holographic CFTs (CFTs with a simple holographic dual- Einstein's gravity) to BCFTs \cite{Hartman:2013mia}, namely large central charge and sparseness of the spectrum of light operators.

In this paper, we 
merge these two lines of research, providing both a purely field theoretic and a gravitational computation for two communicating black holes. The black holes are coupled to a common non-gravitating bath, but from the point of view of one of the black holes, the bath it is coupled to is composed of a gravitating and a non-gravitating part. This is to be contrasted with the original constructions~\cite{Almheiri:2019psy,Almheiri:2019yqk} when the bath (from the point of view of the only black hole) was purely non-gravitating, and the recent work~\cite{Geng:2020fxl} where the bath (from the point of view of one of the two black holes) was purely gravitating.

The field theory setup we consider is a two dimensional CFT living on a strip\footnote{The strip is the time evolution of an interval.} with generically \textit{different} conformal boundary conditions \cite{Cardy:2004hm} on the two edges. 
The dual gravitational setup is a three-dimensional AdS bulk with two two-dimensional KR branes, and with induced black holes living on the branes. 
The asymptotic infinities of these braneworld black holes live on the asymptotic conformal boundary of the three-dimensional bulk, but are separated by a finite distance along the asymptotic boundary. 
This setup is to be contrasted with the wedge holography proposal \cite{Akal:2020wfl} which takes the separation of the two KR branes to zero, and for which the proposed field theory dual is one-dimensional.

We will compute the entanglement entropy of a simple bipartition of the strip in both its gravity description and the field theory description, and show that they agree, 
under the usual
conditions for holographic CFTs, namely vacuum dominance for the correlator of twist operators. In particular, we are not sensitive to microphysics beyond the proposed ansatz for the vacuum sector.
Interestingly, assigning 
mismatched boundary conditions on the two edges of the strip will not only affect the two KR branes but also change the geometry of the bulk dual itself.
The reason for this is that  
mismatched boundary conditions source a gradient for fields on the strip which results in a non-vanishing kinetic energy, therefore changing the bulk geometry.
This insight allows us to generalize the AdS$_{3}$/BCFT$_{2}$ correspondence
for a (holographic) CFT$_{2}$ living on a strip with generic conformal boundary conditions.

To
be more specific, we call the lowest energy eigenstate of the Hamiltonian on the strip the ground state, and denote its energy by $E_\text{gs}$.  Denoting the central charge by $c$, $E_\text{gs}$ is generically larger than $-\frac{c}{12}$, given different boundary conditions on the two ends.
We will distinguish this state from the state corresponding to the identity operator, the vacuum state, which has energy $E = -\frac{c}{12}$.
The difference
between the ground state energy $E_{\text{gs}}$ and the vacuum energy $E = -\frac{c}{12}$ is given by $\Delta_{\text{bcc}}$, the scaling dimension of the so-called boundary condition changing operator.

In the presence of an eternal black hole, the BCFT is in the \textit{thermal field double} (TFD) state \cite{Maldacena:2001kr}
which has a finite temperature 
and the effects of $\Delta_{\text{bcc}}$ are screened.
At zero temperature, on the other hand, we can determine the effect of a non-vanishing $\Delta_{\text{bcc}}$ on the bulk geometry explicitly.
Qualitatively different behaviours arise depending on whether the ground state energy is positive or negative, i.e.~whether $\Delta_{bcc}$ is larger than or smaller than $\frac{c}{12}$.
We find that for a positive ground state energy the bulk geometry is a single sided black hole with mass $M =E_{\text{gs}}= \Delta_{\text{bcc}}-\frac{c}{12}>0$, while for a ground state with negative energy $E_{\text{gs}}=\Delta_{\text{bcc}}-\frac{c}{12}<0$ it is a global AdS$_3$ geometry with a conical defect.
Interestingly, in the second case there appears to be a consistency condition which constraints $\Delta_{bcc}$ to be smaller than $\frac{c}{16}$.
Therefore, the holographic duality suggests that there is a gap $(\frac{c}{16}, \frac{c}{12})$ in the spectrum of $\Delta_{bcc}$.
Moreover, when $\Delta_{bcc}<\frac{c}{16}$, the calculations of the entanglement entropy on the bulk and boundary sides can only be consistently matched if the KR branes have non-positive tension, predicting in turn also non-positive boundary entropies in the BCFT description (note however that  the overall entropy is still positive as it must be).

We emphasize that in a given conformal field theory $\Delta_{bcc}$ can be in principle computed if the boundary conditions are known.
However, $\Delta_{bcc}$ (as a function of the boundary conditions) is not universal but depends on the microscopic details of the theory.
In the gravitational description, the boundary conditions are mapped to the tensions of the branes.
Still, we do not expect to be able to determine $\Delta_{bcc}$ in our bottom-up dual gravitational model explicitly, as it does not contain any information about the microscopic details of a specific boundary theory.
Nonetheless, we can choose $\Delta_{bcc}$ and the brane tensions independently (within the previously mentioned constraints) to match our gravitational computations with the field theory results.

This paper is organized as follows. In Sec.~\ref{sec:review} we review relevant previous work and set up the context for our study. In Sec.~\ref{sec:CFT} we provide the calculations for the entanglement entropy on the CFT side. In Sec.~\ref{sec:gravity} we compute this entanglement entropy on the gravity side and show the agreement with CFT side. We conclude in Sec.~\ref{sec:conclusion} with discussions and possible future directions.

\section{Review of previous work}\label{sec:review}
In this section, we review the relevant background material \cite{Rozali:2019day,Sully:2020pza,Akal:2020wfl,Geng:2020fxl} and set up the context for our calculations. We start with a lightning review of the recently proposed wedge holography \cite{Akal:2020wfl,Miao:2020oey} from the point of view of Karch-Randall braneworlds \cite{Karch:2000gx,Karch:2000ct} and the AdS/CFT correspondence. 
Then we come to the system we want to study in this paper: two communicating black holes. This system was first studied in higher dimensions using holographic tools in \cite{Geng:2020fxl}, an approach we review in detail.
We then discuss the BCFT calculation of entanglement entropy outlined in \cite{Sully:2020pza}. 

\subsection{Wedge holography}
The
AdS/CFT correspondence \cite{Maldacena:1997re,Gubser:1998bc,Witten:1998qj} states that quantum gravity in an asymptotically AdS$_{d+1}$ space is dual to a $d$-dimensional conformal field theory living on its asymptotic boundary. This asymptotic boundary is usually called the \textit{conformal boundary} as the isometry group of AdS$_{d+1}$ is the conformal group of a $d$-dimensional conformal field theory. 

The observation that an AdS$_{d+1}$ space can be foliated by an AdS$_d$ space (see e.g.~\cite{Karch:2000ct}) makes this correspondence even more powerful, and allows a way to relate descriptions in spacetimes with a difference of two (spacelike) dimensions.
This realization has been fruitful in diverse setups, e.g.~in the context of asymptotic symmetries of AdS space~\cite{Mishra:2017zan}, and
in the context of black hole information paradox~\cite{Almheiri:2019hni}. A dictionary for this \textit{codimension two holography} duality was given in~\cite{Akal:2020wfl,Miao:2020oey}, using which observables of interest were calculated there.\footnote{It is interesting to notice that \cite{Miao:2021ual} recently generalized these studies to codimension-n holography.} 

\begin{figure}[h]
\begin{centering}
\subfloat[Wedge Holography\label{wedgeholography}]
{
\begin{tikzpicture}[scale=1.4]
\draw[-,very thick,black!40] (-2,0) to (0,0);
\draw[-,very thick,black!40] (0,0) to (2,0);

\draw[-,very thick] (0,0) to (2,-0.6);

\draw[-,very thick] (0,0) to (-2,-1.66);

%\draw[pattern=north west lines,pattern color=blue!200,draw=none] (0,0) to (-2,-1.66) to (-2,0) to (0,0);
%\draw[pattern=north west lines,pattern color=blue!200,draw=none] (0,0) to (2,0) to (2,-0.2) to (0,0);
\draw[fill=gray, draw=none, fill opacity = 0.1] (0,0) to (-2,-1.66) to (-2,0) to (0,0);
\draw[fill=gray, draw=none, fill opacity = 0.1] (0,0) to (2,0) to (2,-0.6) to (0,0);
%violet!50

\draw[-,dashed,color=blue] (0,0) to (-1.5,-2); 
\draw[-,dashed,color=blue] (0,0) to (0,-2.15);
\draw[-,dashed,color=blue] (0,0) to (1.5,-2); 
\draw[-,dashed,color=blue] (0,0) to (2,-1.5); 

\draw[-] (-0.75,0) arc (180:217.5:0.8);
\node at (-0.85,-0.25) {$\theta_1$};

\draw[-] (.85,0) arc (0:-4.95:2.75);
\node at (1.2,-0.2) {$\theta_2$};

\draw[-,dashed,color=red, thick] (1,-0.3) arc (-2.25:-90-52.5-10:1.005);
\draw[-,dashed,color=red, thick] (1.5,-0.5) arc (-2.25:-90-52.5-11.5:1.5);
\node at (0,0) {\textcolor{red}{$\bullet$}};
\node at (0,0) {\textcolor{black}{$\circ$}};

\draw[->,thick,color=black] (0.0,0.2) to (1,0.2);
\node at (0.5,0.4)
{\textcolor{black}{$w$}};

\draw[->,thick,color=black] (-2.2,0) to (-2.2,-1);
\node at (-2.4,-0.5)
{\textcolor{black}{$z$}};

%\node at (-1.25,-1.5) [rotate=50][scale=0.7]
%{\textcolor{black}{$\mu = \text{const}$}};

%\node at (0.25,-1.5) [rotate=20][scale=0.7]
%{\textcolor{black}{$u = \text{const}$}};

\draw[->] (-1.0,-1.85) arc (210:265:0.8);
\node at (-0.7,-1.95) {$\mu$};

\draw[->] (1.0,-1.2) to (1.5, -1.75);
\node at (1.45,-1.45) {$u$};

\end{tikzpicture}
}
\hspace{1.0cm}
\subfloat[UV Regularized Wedge Holography \label{UVWedge}]
{

\begin{tikzpicture}[scale=1.4]
\draw[-,very thick,black!40] (-2,0) to (0,0);
\draw[-,very thick,black!40] (0,0) to (3,0);

\draw[-,very thick] (1,0) to (2+1,-0.5);

\draw[-,very thick] (0,0) to (-2,-1.5);

%\draw[pattern=north west lines,pattern color=blue!200,draw=none] (0,0) to (-2,-1.5) to (-2,0) to (0,0);
%\draw[pattern=north west lines,pattern color=blue!200,draw=none] (1,0) to (3,0) to (3,-0.5) to (1,0);
\draw[fill=gray, draw=none, fill opacity = 0.1] (0,0) to (-2,-1.5) to (-2,0) to (0,0);

\draw[fill=gray, draw=none, fill opacity = 0.1] (1,0) to (3,0) to (3,-0.5) to (1,0);
%violet!50

\draw[-] (-0.75,0) arc (180:217.5:0.75);
\node at (-1.0,-0.35) {$\theta_1$};

\draw[-] (1.5,0) arc (0:-5.25:1.5);
\node at (2.3,-0.17) {$\theta_2$};

\node at (0,0) {\textcolor{red}{$\bullet$}};
\node at (0,0) {\textcolor{black}{$\circ$}};

\node at (1,0) {\textcolor{red}{$\bullet$}};
\node at (1,0) {\textcolor{black}{$\circ$}};

\draw[-,thick,color=black!50] (0,0.05) to (0,0.25); 
\draw[-,thick,color=black!50] (1,0.05) to (1,0.25);
\draw[<-,thick,color=black!50] (0,0.15) to (0.3,0.15);
\draw[->,thick,color=black!50] (0.7,0.15) to (1,0.15);
\node at (0.5,0.2)
{\textcolor{black}{$L$}};

\node at (0.5,-2.15)
{\textcolor{black}{}};

\end{tikzpicture}
}
\caption{\small{\textit{(a) An illustration of wedge holography: the bulk geometry is a part of empty AdS$_{d+1}$ between two end-of-world Karch-Randall branes (black lines), meeting each other on the conformal boundary of AdS$_{d+1}$ at their common boundary (the red dot). The bulk shaded regions behind the branes are removed, leaving only a wedge in the bulk. Wedge holography states that the gravitational physics in the bulk wedge is dual to a $(d-1)$-dimensional conformal field theory living on the red dot (the defect). (b) The wedge holography proposal can be UV regularized 
by considering a BCFT$_d$ with two boundaries (the two red dots). By sending the separation $L$ between the two boundaries to zero, one goes back to the wedge.}}}
\label{allbcft2branes}
\end{centering}
\end{figure}
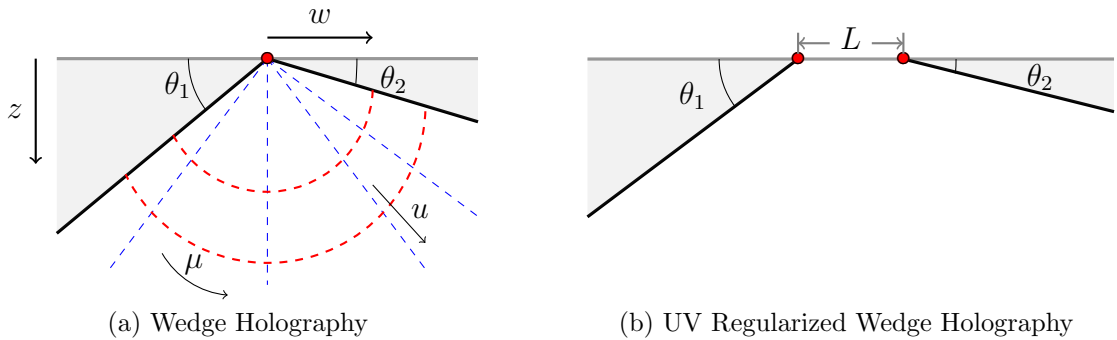

Consider such a foliation of AdS$_{d+1}$ space by AdS$_d$ space (represented by dashed blue lines in Fig.~\ref{wedgeholography}). These slices meet each other on the AdS$_{d+1}$ conformal boundary at the defect (the red dot in Fig.~\ref{wedgeholography}), which is the conformal boundary for each AdS$_{d}$ slice. To see this, it is simplest to start from the standard Poincar\'{e} patch of the AdS$_{d+1}$, with metric
\begin{equation}\label{eq:AdSPoincare}
    ds^2=\frac{-dt^2+dw^2+d\vec{x}^2+dz^2}{z^2}\,,\quad z>0\, ,
\end{equation}
and perform the coordinate transformation
\begin{equation}
    w=u\cos\mu\,,\quad z=u\sin\mu\,.
\end{equation}
This yields the metric
\begin{equation}
    ds^2=\frac{1}{\sin^{2}\mu}\left(\frac{-dt^2+d\vec{x}^2+du^2}{u^2}+d\mu^2\right)\,, \qquad 0 < \mu < \pi\,,\; u>0\:.
\end{equation}
Each constant $\mu$ slice (dashed blue lines in Fig.~\ref{wedgeholography}) is a copy of AdS$_{d}$ with conformal boundary at $u=0$. Then we can generate a wedge in the $(d+1)$-dimensional bulk by putting two positive tension branes along the $\mu=\theta_1$ and $\mu=\pi-\theta_2$ slices (the two black lines in Fig.~\ref{wedgeholography}). These are called Karch-Randall branes \cite{Karch:2000ct,Karch:2000gx}, and their tensions are determined by solving the Israel's junction conditions \cite{Kraus:1999it,Takayanagi:2011zk,Fujita:2011fp}. In the bulk, the two KR branes remove the portion of the spacetime behind them (gray region in Fig.~\ref{wedgeholography}),\footnote{More precisely, 
in string theory constructions, the KR branes are the locations where extra compact dimensions degenerate \cite{Chan:2000ms} and so are also called the end-of-world (EOW) branes. See \cite{Uhlemann:2021nhu} for a recent string theory realization of the wedge holography and communicating black holes.} giving us a wedge as the bulk.

The AdS/CFT correspondence for braneworlds \cite{ArkaniHamed:2000ds} tells us that quantum gravity in the $(d+1)$-dimensional wedge is dual to UV cutoff CFTs living on the two KR branes, coupled to dynamical gravity. Hence, this lower dimensional dual description is also a theory of quantum gravity, with two sectors each living on an AdS$_{d}$ space. Using the standard AdS/CFT correspondence \cite{Maldacena:1997re,Gubser:1998bc,Witten:1998qj}, we can further dualize this $d$-dimensional description to a $(d-1)$-dimensional CFT living on the common conformal boundary of the branes, the defect itself. In principle, this $(d-1)$-dimensional description has two sectors corresponding to the two KR branes.

This co-dimension two wedge holography can be extrapolated from the standard AdS/BCFT duality by first regularizing the UV, and then taking the regulator to zero~\cite{Akal:2020wfl}. 
The most straightforward regularization separates the two Karch-Randall branes by a distance $L$ (see Fig.~\ref{UVWedge}) on the conformal boundary.
In the Poincar\'{e} patch of the bulk AdS$_{d+1}$, the two branes have the embedding equations:
\begin{align}
&\text{Left Brane:}&w=u\cos\theta_1\,, \quad z=u\sin\theta_1\,,\nonumber\\
&\text{Right Brane:}&w=-L-u\cos\theta_2\,, \quad z=u\sin\theta_2\,.
\label{eq:braneembedding}
\end{align}
The dual description is a $d$-dimensional CFT living on the conformal boundary of the regulated wedge, with two parallel boundaries (the two red dots on Fig.~\ref{UVWedge}). 
We will study the entanglement entropies associated with a bipartition which cuts this UV-regularizing interval into two, along a $(d-1)$-dimensional sub-manifold parallel to the two defects.

\subsection{Two communicating black holes}
In \cite{Geng:2020fxl}, a model of two communicating black holes was constructed by placing two KR branes into an AdS$_{d+1}$ black string with line element
\begin{equation}\label{eq:blackstring}
    ds^2=\frac{1}{\sin^{2}\mu}\left(\frac{-h(u)\,dt^2+ h(u)^{-1}\,du^2+d\vec{x}^2}{u^2}+d\mu^2\right),\quad h(u)=1-\frac{u^{d-1}}{u_{h}^{d-1}} \,,
\end{equation}
where $u_h$ is the location of the horizon. Each of the branes at $\mu=\theta_{1}$ and $\mu=\pi-\theta_{2}$ supports an AdS$_{d}$ planar black hole, which in turn are coupled at their common conformal boundary and are in thermal equilibrium with each other.
Although in equilibrium, they exchange information in the form of Hawking radiation, emitting quanta and swallowing quanta from the other. From the viewpoint of each of the black holes, the bath to which it is coupled is gravitating.

This communication between the black holes can be characterized by considering the entanglement entropy of an internal bipartition. 
This internal bipartition makes use of the fact that the defect Hilbert space can be factorized into two sectors roughly corresponding to the two black holes. This entanglement entropy was referred to as the L/R entropy in Ref.~\cite{Geng:2020fxl} (see details therein).
To study the time evolution of the entanglement entropy, the time-translation of the system has to be chosen such that the two disconnected components of the system (corresponding to the two asymptotic boundaries in the maximally extend bulk geometry) evolve in the same direction.\footnote{In 
the bulk we
considered an eternal AdS$_{d+1}$ black string whose field theory dual is known to be in a thermofield double state \cite{Maldacena:2001kr}.
We applied a general trick used when studying time evolution of the thermofield double state. By choosing the overall Hamiltonian $H$ to be $H_1 + H_2$ rather than $H_1 - H_2$, there is a non-trivial time evolution.}
As noticed in \cite{Geng:2020fxl}, this entanglement entropy can be studied using the Ryu-Takayanagi (RT) surface in the $(d+1)$-dimensional bulk, and the RT surface has a phase transition between a connected phase and a disconnected phase. This phase transition gives a Page curve for the L/R entropy, which is consistent with the unitarity of the time evolution we chose.

However, it is not easy to give a calculable field theory model of this L/R entropy, since the black string geometry does not have a natural UV regularization preserving the conformal symmetry on the defect.
In particular, in Eq.~\eqref{eq:blackstring} the translational symmetry of the AdS metric in Eq.~\eqref{eq:AdSPoincare} along the $w$-coordinate is broken, and hence it is not possible anymore to embed two branes with a separation as in Eq.~\eqref{eq:braneembedding} in a consistent way.
Therefore, instead of a black string, we will consider an eternal AdS$_{3}$ BTZ black hole \cite{Banados:1992wn} with line element
\begin{equation}
    ds^2=-\frac{h(z)}{z^2}dt^2+\frac{dz^2}{h(z)z^2}+\frac{dx^2}{z^2}\,, \quad h(z)=1-\frac{z^2}{z_{h}^2} \,,
\end{equation}
where $z_h$ is the location of the horizon. Contrary to the higher-dimensional case, in three dimensions KR branes can be consistently embedded in such a black hole background, where the brane-localized action is just a tension term. Moreover, now the two KR branes can be separated by a distance on the conformal boundary.
Each brane contains a two-dimensional black hole induced from the bulk BTZ, a construction discussed in more detail later (Sec.~\ref{sec:gravity}). Thus, we have a conformal symmetry-preserving, UV-regularized
description of two communicating black holes. The field theory description consists of two BCFT$_{2}$s, each with two boundaries. These two BCFTs live on the conformal boundary of the two copies of the system (since the maximal extension of the Penrose diagram corresponding to the eternal black holes has two conformal boundaries). On the time reflection-symmetric slice, these two BCFTs are in a thermofield double (TFD) state.

Based on this observation, we will see that we can have a field theory computation of the L/R entropy (see Sec.~\ref{sec:CFT}) which matches the gravity side computation in Sec.~\ref{sec:gravity}. This provides an explicitly calculable field theory model for two communicating black holes. 
Furthermore, it provides another consistency check for the AdS/CFT correspondence and its possible deformations.

\subsection{BCFT calculations of entanglement entropy}\label{sec:reviewBCFTEE}

In this section, we review the computation of entanglement entropy in two-dimensional BCFTs, mostly following \cite{Sully:2020pza}, and clarify some subtleties. We start from the basic concepts of computing the entanglement entropy using the replica trick. We list useful results for our computations in the following sections and state our assumptions. 
\subsubsection{Boundary state at zero temperature}
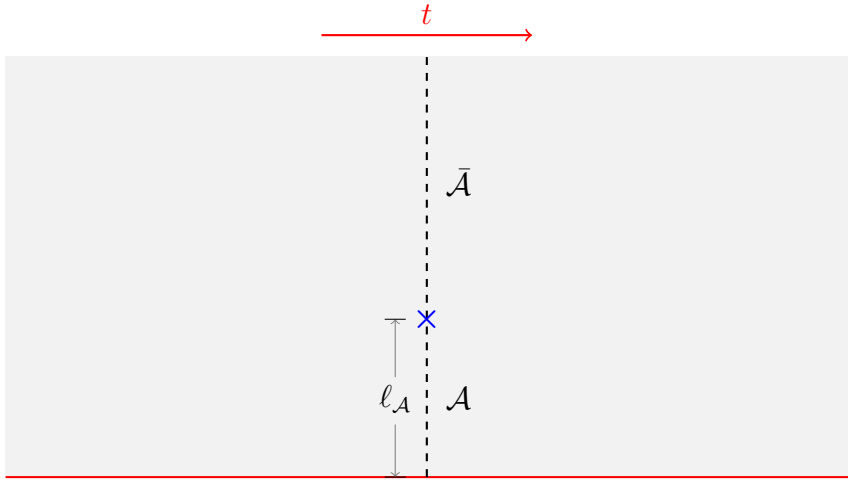
\begin{figure}[h]
\begin{centering}
{
\begin{tikzpicture}[scale=1.4]
\draw[-, thick,red] (-4,0) to (4,0);

%\draw[pattern=north west lines,pattern color=gray!200,draw=none] (-4,0) to (4,0) to (4,4) to (-4,4);
\draw[fill=gray, draw=none, fill opacity = 0.1] (-4,0) to (4,0) to (4,4) to (-4,4) to (-4,0);
\draw[->, thick,red] (-1,4.2) to (1,4.2);
\node at (0,4.4)
{\textcolor{red}{$t$}};
%\draw[-,dashed,color=black!50!lime,very thick] (0,0) to (0,4);
\draw[-,dashed,color=black, thick] (0,0) to (0,4);
\node at (0,1.5) {\textcolor{blue}{$\cross$}};
\node at (0.3,0.75)
{\textcolor{black}{$\mathcal{A}$}};
\node at (0.3,2.8)
{\textcolor{black}{$\bar{\mathcal{A}}$}};
\node at (-0.3,0.75)
{\textcolor{black}{$\ell_\mathcal{A}$}};
\draw[->, thin,gray] (-0.3,0.95) to (-0.3,1.5);
\draw[->, thin,gray] (-0.3,0.5) to (-0.3,0);
\draw[-, thin,black] (-0.4,1.5) to (-0.2,1.5);
\draw[-, thin,black] (-0.4,0) to (-0.2,0);
\end{tikzpicture}
}
\caption{\small{\textit{We consider a two dimensional boundary conformal field theory living on the upper half plane (UHP). The bulk CFT is in the vacuum state, the boundary is specified by the horizontal axis (in red) and the time direction is along the horizontal direction. We consider a constant time slice (dashed black vertical line) which defines the state we are looking at. For this state, we want to compute the entanglement entropy of the region $\mathcal{A}$ (between the blue cross and the boundary) with its complement $\bar{\mathcal{A}}$. }}}
\label{demonEE}
\end{centering}
\end{figure}

For the sake of convenience, we consider an Euclidean BCFT$_2$ and firstly assume that the bulk CFT is in the vacuum state. In this case, the CFT is living on an upper half plane and the boundary condition on the real axis preserves half of the conformal invariance in the bulk \cite{Cardy:2004hm}. The Euclidean time $t$ is going along the horizontal axis (see Fig.~\ref{demonEE}). We want to compute the entanglement entropy of the subsystem $\mathcal{A}$ in Fig.~\ref{demonEE}. Suppose that the reduced density matrix of this subsystem is $\rho_{\mathcal{A}}$, then the entanglement entropy can be computed by taking the limit $n\rightarrow1$ of its n-th Renyi entropy
\begin{equation}
    S_{\mathcal{A}}^{n}=-\frac{1}{n-1}\ln\Tr(\rho_{\mathcal{A}}^n) \,. \label{eq:defintion}
\end{equation}
The trace $\Tr(\rho_{\mathcal{A}}^n)$ in the above formula can be computed by the standard replica trick \cite{Calabrese:2009qy}. It is equivalent to the one-point function of a twist operator $\Phi_{n}(z,\bar{z})$ inserted at the point that separates $\mathcal{A}$ and its complement $\bar{\mathcal{A}}$ (the blue cross in Fig.~\ref{demonEE}). This twist operator creates a branch cut on the upper half plane (UHP) from the blue cross to infinity along the dashed black line in Fig.~\ref{demonEE}.

The twist operator is a primary operator \cite{Calabrese:2009qy} with conformal dimensions
\begin{equation}
    h_{n}=\bar{h}_{n}=\frac{c}{24}\left(n-\frac{1}{n}\right).
\end{equation}
Therefore, we have the following formula for the entanglement entropy of the subsystem $\mathcal{A}$:
\begin{equation}
    S_{\mathcal{A}}=\lim_{n\rightarrow1}\frac{1}{1-n}\ln\langle\Phi_{n}(z,\bar{z})\rangle_{\text{UHP}} \,, \label{eq:von}
\end{equation}
where $(z,\bar{z})$ is the location of the boundary of $\mathcal{A}$ (the blue cross in Fig.~\ref{demonEE}). This one-point function is fixed by conformal symmetry and can be computed using the doubling trick \cite{Sully:2020pza} to map the BCFT on the UHP to a chiral CFT on the whole complex plane $\mathbb{C}$. Doing this trick, the one-point function for $\Phi_{n}(z,\bar{z})$ is equivalent to a two-point function of a chiral primary field $\Phi_{n}(z)$ (with conformal dimensions $h_n=\frac{c}{24}\left(n-\frac{1}{n}\right),\,\bar{h}_{n}=0$) where one $\Phi_n$ is inserted at the blue cross and the other $\Phi_n$ at its mirror point on the lower half plane,
\begin{equation}
        \langle\Phi_{n}(z,\bar{z})\rangle_{\text{UHP}}=\langle\Phi_{n}(z)\Phi_{n}(z^*)\rangle_{\mathbb{C}}=\frac{\mathcal{A}_{\Phi_{n}}^{b}}{|z-z^{*}|^{2h_{n}}} \,,\label{eq:1pUHP}
\end{equation}
where $\mathcal{A}_{\Phi_{n}}^{b}$ is a normalization constant that will be fixed soon \cite{Sully:2020pza}.

This trick is equivalent to a boundary operator expansion (BOE) (see \cite{Cardy:1996xt,Wu:2020mxj} for explicit calculations of BOEs in simple models) which states that on the UHP any operator $\mathcal{O}_{i}(z,\bar{z})$ can be expanded as a series of boundary operators $\hat{O}_{I}(x)$,
\begin{equation}
    \mathcal{O}_{i}(z,\bar{z})=\sum_{J}\frac{\mathcal{B}_{i}^{bJ}}{(2y)^{\Delta_{i}-\Delta_{J}}}\tilde{\mathcal{C}}[y,\partial_{x}]\hat{\mathcal{O}}_{J}(x) \,.\label{eq:BOE}
\end{equation}
Here $\Delta_{i}$ is the conformal weight of $\mathcal{O}_{i}$ ($h_{i}=\bar{h}_{i}=\frac{\Delta_{i}}{2}$), $\Delta_{J}$ is that of the boundary primary operator $\hat{\mathcal{O}}_{J}(x)$ and we use the complex coordinate $z=x+iy$ ($y>0$) on the UHP. The coefficients $\mathcal{B}_{i}^{bJ}$ are called the BOE coefficients and they are determined by the boundary condition and the structure of the parent CFT. The boundary primary operator is normalized as
\begin{equation}
    \langle\hat{O}_{I}(x_{I})\hat{O}_{J}(x_{J})\rangle=\frac{G_{IJ}}{|x_{I}-x_{J}|^{2\Delta_{I}}} \,.
\end{equation}
By comparing Eq.~\eqref{eq:1pUHP} with the vacuum expectation value of Eq.~\eqref{eq:BOE}, we see that the only boundary operator in the BOE that appears in $\left<\Phi_n\right>_{\text{UHP}}$ is the identity operator $\mathbf{1}$, and 
\begin{equation}
    \mathcal{A}_{\Phi_{n}}^{b}=\mathcal{B}_{\Phi_{n}\mathbf{1}}^{b} \,.
\end{equation}
Here the index $J$ in $B_{i}^{bJ}$ is lowered using $G_{IJ}$. Now we have an exact result for the entanglement entropy of $\mathcal{A}$:
\begin{equation}
    S_{\mathcal{A}}=\frac{c}{6}\ln(\frac{2\ell_{\mathcal{A}}}{\epsilon})+\ln(g_{b}) \,,
\end{equation}
where $\ell_{\mathcal{A}} = (z-\bar{z})/2i$ is the length of the subsystem $\mathcal{A}$ and we have defined the boundary entropy term $\ln(g_{b})$ through the regularization 
\begin{equation}
    \ln(g_{b})-\frac{c}{6}\ln(\epsilon)=\lim_{n\rightarrow1}\frac{1}{1-n}\ln(B_{\Phi_{n}1}^{b})\, .
    \label{eq:bdyentropy}
\end{equation}
Here $\epsilon$ is a UV cutoff. We note that this result matches the holographic computation in \cite{Takayanagi:2011zk} and is of a universal form for any BCFT with one boundary such as in Fig.~\ref{demonEE}.

\subsubsection{Thermal field double state}\label{sec:ReviewTFD}
\begin{figure}[h]
\begin{centering}
\subfloat[\textit{The thermofield double state and its time evolution}\label{TFDpre1}]
{
\begin{tikzpicture}[scale=0.9]
\draw[-,color=red, thick] (-1,0) arc (1:361:-1);
\draw[-,dashed,color=black, thick] (-3.5,0) to (-1,0);
\draw[-,dashed,color=black, thick] (1,0) to (3.5,0);
\node at (-2,-0.4)
{\textcolor{black}{BCFT$_L$}};
\node at (2,-0.4)
{\textcolor{black}{BCFT$_R$}};
%\draw[-,dashed,color=darkpastelgreen, thick](0.705,0.705) to (1.41+0.705,1.41+0.705);
%\draw[-,dashed,color=darkpastelgreen, thick](-0.705,0.705) to (-1.41-0.705,1.41+0.705);
\draw[-,dashed,color=darkpastelgreen, thick](0.705+0.5,0.705+0.5) to (1.41+0.705,1.41+0.705);
\draw[-,dashed,color=darkpastelgreen, thick](-0.705-0.5,0.705+0.5) to (-1.41-0.705,1.41+0.705);
\draw[-,color=darkpastelgreen, thick](0.705,0.705) to (0.705+0.5,0.705+0.5);
\draw[-,color=darkpastelgreen, thick](-0.705,0.705) to (-0.705-0.5,0.705+0.5);

\draw[<-,, color=red, thin] (0.7,0.4) arc (25:-25:1);
\node at (0.45,0.2) {\textcolor{black}{$t_R$}};

\draw[<-,, color=red, thin] (-0.7,0.4) arc (-25:25:-1);
\node at (-0.45,0.2) {\textcolor{black}{$t_L$}};

\node at (0.705+0.5,0.705+0.5) {\textcolor{blue}{$+$}};
\node at (-0.705-0.5,0.705+0.5) {\textcolor{blue}{$+$}};
\node at (0.705,0.705+0.5) {\textcolor{black}{$\mathcal{A}_R$}};
\node at (-0.705,0.705+0.5) {\textcolor{black}{$\mathcal{A}_L$}};
\node at (0.705+0.75,0.705+0.5+0.75) {\textcolor{black}{$\bar{\mathcal{A}}_R$}};
\node at (-0.705-0.75,0.705+0.5+0.75) {\textcolor{black}{$\bar{\mathcal{A}}_L$}};
\end{tikzpicture}
}
\hspace{0.5cm}
\subfloat[\textit{Conformal mapping of the region on left to a UHP} \label{TFDmap1}]
{
\begin{tikzpicture}[scale=0.9]

\draw[fill=gray, draw=none, fill opacity = 0.1] (-4,0) to (4,0) to (4,3) to (-4,3);
\draw[-, thick,red] (-4,0) to (4,0);

\node at (1.5,1.5) {\textcolor{blue}{$\cross$}};
\node at (-1.5,1.5) {\textcolor{blue}{$\cross$}};
\draw[-,dashed,darkpastelgreen] (-1.5,1.5) to (1.5,1.5);
\node at (-1.5,1.9) {\textcolor{black}{\emph{$\bar{\Phi}_{n}$}}};
\node at (1.5,1.9) {\textcolor{black}{\emph{$\Phi_{n}$}}};
\node at (0,2) {\textcolor{black}{\emph{$\bullet$}}};
\node at (0,2.4) {\textcolor{black}{\emph{$\left(0,\frac{1}{2}\right)$}}};
\end{tikzpicture}

}
\caption{\small{\textit{a) The two Euclidean BCFTs L and R in the TFD state: time evolution is rotation with respect to the origin, hence the red circle is the time evolution of the boundary. The two black dashed lines define the zero time slice.
Under the chosen time evolution, L and R evolve clockwise and counter-clockwise respectively, as indicated.
We are interested in the entanglement entropy of the subsystem $\mathcal{A}=\mathcal{A}_{L}\cup\mathcal{A}_{R}$ corresponding to the solid green line segments. 
b) The UHP which results from the conformal mapping of the region outside of the red circle in Fig.~\ref{TFDpre1}. The circular boundary is mapped to the real axis and infinity is mapped to $\left(0,\frac{1}{2}\right)$. The location of twist operators is mapped to the two blue crosses, separated in the horizontal direction. The branch cut is mapped (and deformed) to the dashed green line connecting the two operators.}}}
\label{TFDall}
\end{centering}
\end{figure}
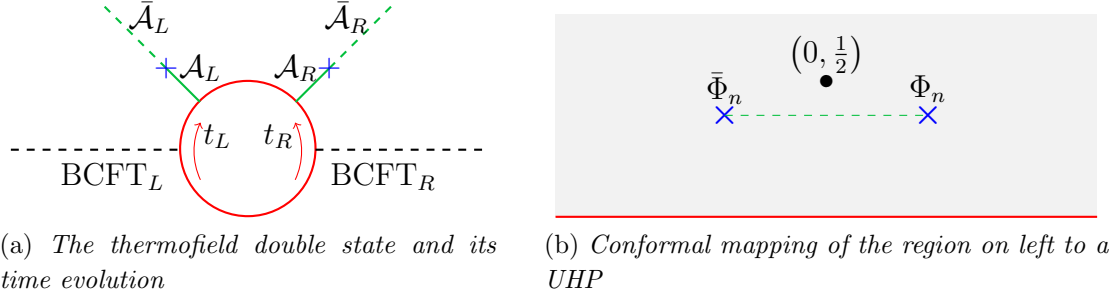

We now consider the thermal field double (TFD) state 
of two BCFTs.
We will label the two BCFTs as L and R respectively (not to be confused with the L/R entropy of the previous sections, the meaning should be clear by context). 
We can prepare the TFD state using an Euclidean path integral. In the gravitational picture, it corresponds to the Hartle-Hawking state in the maximally extended two-sided geometry.
In this Euclidean signature, the time direction is periodic, and a time evolution for the L and R parts can be chosen that makes the TFD state evolve non-trivially (see Fig.~\ref{TFDpre1}).  We would like to calculate the entanglement entropy of the subsystem $\mathcal{A}_L \cup \mathcal{A}_R$. As in the previous subsection, the replica trick tells us that the question reduces to the computation of the two-point function of a twist operator $\Phi_n$ and an anti-twist operator $\bar{\Phi}_n$ inserted respectively at the points which separate $\mathcal{A}_L (\mathcal{A}_R)$ and its complement $\bar{\mathcal{A}}_L (\bar{\mathcal{A}}_R)$ (blue crosses in Fig.~\ref{TFDpre1}). Denoting the location of the crosses by $w_L$ and $w_R$, the relevant two-point function is
\begin{equation}
    \langle\Phi_{n}(w_{R},\bar{w}_{R})\bar{\Phi}_{n}(w_{L},\bar{w}_{L})\rangle \,.
\end{equation}
The twist operator creates a branch cut from the right blue cross to infinity along the right dashed green line and the anti-twist operator creates a branch cut from the left blue cross to infinity along the left dashed green line.\footnote{The twist operator and the anti-twist operator are related to each other by a reflection (parity only in one spatial direction) transformation.} 
To compute the two-point function of the twist and anti-twist operators we can map the configuration to the upper half plane in z coordinates using the conformal transformation
\begin{equation}
    w=\frac{1}{z-\frac{i}{2}}-i\label{eq:TFDmap1} \,.
\end{equation}
This transformation maps the boundary circle in $w$ coordinates (red circle in Fig.~\ref{TFDpre1}) to the real axis in $z$ and the infinity in $w$ to $z = i/2$. 
The blue crosses are mapped to two point on the UHP. The branch cuts from the blue crosses to infinity can be deformed to a horizontal branch cut connecting them (the dashed green line in Fig.~\ref{TFDmap1}) on the UHP.

As a result, we have to compute the two-point function of a twist operator and an anti-twist operator on the UHP:
\begin{equation}
    \langle\bar{\Phi}_{n}(z_{L},\bar{z}_{L})\Phi_{n}(z_{R},\bar{z}_{R})\rangle_{\text{UHP}} \,.
\end{equation}

When there is no boundary, this is just the two-point function of primary operators which is totally fixed by kinematics:
\begin{equation}
    \langle\bar{\Phi}_{n}(z_{L},\bar{z}_{L})\Phi_{n}(z_{R},\bar{z}_{R})\rangle_{\mathbb{C}}=\frac{\epsilon^{2d_{n}}}{|z_{L}-z_{R}|^{2d_{n}}} \,,
\end{equation}
where the conformal weight $d_{n}$ equals to $2h_{n}=2\bar{h}_{n}=\frac{c}{12}(n-\frac{1}{n})$ and we have set the normalization to be $\epsilon^{2d_{n}}$, (where $\epsilon$ is a UV cutoff) for reasons that will be clear when we compute the entanglement entropy.

However, when we have a boundary, 
there is an ambiguity. To see this, first note that, away from the boundary, the BCFT has the same local structure as its parent CFT. This means that the operator product expansion (OPE) for two bulk operators is the same as in the parent CFT:
\begin{equation}
    \mathcal{O}_{i}(z_{1},\bar{z}_{1})\mathcal{O}_{j}(z_{2},\bar{z}_{2})=\sum_{k}\frac{\hat{\mathcal{C}}_{ij}^{k}}{|z_{1}-z_{2}|^{\Delta_{i}+\Delta_{j}-\Delta_{k}}}C_{\Delta_{i}\Delta_{j}\Delta_{k}}[z_{12},\bar{z}_{12},\partial_{2},\bar{\partial}_{2}]\mathcal{O}_{k}(z_{2},\bar{z}_{2}) \,,\label{eq:OPE}
\end{equation}
for bulk primaries $\mathcal{O}_{k}$ 
and OPE coefficients $\mathcal{\hat{C}}_{ij}^{k}$. 
To compute our two-point function, we can use the OPE and compute the one-point functions of the primaries $\mathcal{O}_k$ appearing in the product, as per Eq.~\eqref{eq:1pUHP} and Eq.~\eqref{eq:BOE}.
We call this the bulk channel.
Alternatively, we can perform a BOE (Eq.~\eqref{eq:BOE}) first and then compute the (normalized) boundary two-point functions. We call this the boundary channel.

Consistency of the parent CFT and the boundary conditions tell us that these two channels should give the same answer.
This consistency requirement allows us to simplify the computation of 2-point function in the case of a holographic BCFT, which we now define.

First recall that for a holographic (two-dimensional) CFT we have \cite{Hartman:2013mia}:
\begin{itemize}
    \item The central charge $c$ is large.
    \item The spectrum of light operators, of conformal dimension $\mathcal{O}(c^{0})$,\footnote{In this paper, the $\mathcal{O}(\dots)$ symbol means "of the order $\dots$".} is sparse.
\end{itemize}
The first condition tells us that in the OPE the operators of conformal dimension $\mathcal{O}(c)$ or higher, which we refer to as \textit{heavy} operators, are suppressed (for technical details see \cite{Sully:2020pza,Hartman:2013mia,Faulkner:2013ana}). The second condition tells us that the total contributions to the OPE from the light operators is a multiplication of the contribution of the identity operator. These conditions greatly simplify the OPE for holographic CFTs because we only have to look at the fusion to the identity operator.

For holographic BCFTs, the two assumptions are naturally generalized to boundary operators \cite{Sully:2020pza}, but some subtleties appear. First let us consider the bulk channel. In the OPE, one-point functions of purely bulk operators are now generically nonzero, and may compete with the suppressed OPE contribution from the heavy operators when we compute a bulk two-point function.
However, this happens only when we are close enough to the boundary, where the position-dependent part of the heavy operator one-point function (calculated using the doubling trick in the previous subsection) is not small.

Now consider the boundary channel. The contribution of a boundary operator of dimension $\Delta_J$ to the BOE of $\Phi$ scales as $y^{\Delta_J}$ (see Eq.~\eqref{eq:BOE}). For large $y$, the contribution from heavy operators is not suppressed anymore. Therefore, if we want to only look at the contribution of the boundary identity operator, we must not be too far from the boundary. 

In summary, the simplification on the calculation of the bulk two-point function from using the conditions of holographic BCFTs is useful in the bulk channel only when far away from the boundary, and in the boundary channel only when close to the boundary. Consistency of the BCFT requires that the bulk channel result must equal the boundary channel result.

Practically speaking, when we compute the bulk two-point function for a given pair of insertions, we examine the identity block in both channels and simply choose the larger one. This is justified because each contribution to the two-point function from different blocks is positive
and taking all corrections into account should give the same result for both channels. 
Therefore, the larger of the two identity blocks is a better approximation.
In other words, in the other channel with a smaller identity block contribution neglecting everything except for the identity block gives rise to a larger error.
This takes into account the subtleties due to the appearance of the boundary.

As a result, for a holographic BCFT on an UHP we have the two-point function
\begin{equation}
    \langle\bar{\Phi}_{n}(z_{L},\bar{z}_{L})\Phi_{n}(z_{R},\bar{z}_{R})\rangle_{\text{UHP}}=\max\left(\frac{\epsilon^{2d_{n}}}{|z_{L}-z_{R}|^{2d_{n}}},\frac{\mathcal{B}_{\Phi_{n}1}^{b}\mathcal{B}_{\Phi_{n}1}^{b}}{|z_{L}-z_{L}^{*}|^{d_{n}}|z_{R}-z_{R}^{*}|^{d_{n}}}\right) \, ,
\end{equation}
where the first (second) term is the contribution from bulk (boundary) channel.
We can immediately deduce the entanglement entropy, but we defer details to the next section where we consider the more general case of two boundaries.
We briefly note, however, that the maximization prescription for the two-point function gives a \textit{minimization} prescription for the entanglement entropy (due to the minus sign in Eq.~\eqref{eq:defintion}), consistent with the Ryu-Takayanagi prescription in holographic computations \cite{Ryu:2006bv,Ryu:2006ef}.

\section{The CFT computation}\label{sec:CFT}
In this section, we discuss the two-dimensional CFT on a strip, and calculate the entanglement entropy of a subsystem.
Before moving to the finite temperature case, we consider the zero temperature case first.

\subsection{CFT with two boundaries, at zero temperature}\label{sec:Vacuum}

We consider a two dimensional Euclidean CFT, living on a strip, with \textit{different} boundary conditions along the two boundaries. We denote these boundary conditions as $a$ and $b$ respectively.

\begin{figure}[h]
\begin{centering}
{
\subfloat[\textit{CFT on a strip with boundary conditions a, b.}\label{vacuum}]
{
\begin{tikzpicture}[scale=0.5]
\draw[-, thick,red!40] (-4,0) to (4,0);
\draw[-, thick,red!40] (-4,4) to (4,4);

\draw[fill=gray, draw=none, fill opacity = 0.1] (-4,0) to (4,0) to (4,4) to (-4,4);
\draw[-,thin,black,dashed] (0,0) to (0,4);
\node at (0,1.5){\textcolor{darkpastelgreen}{$\cross$}};
\draw[->,thin,red] (-1,4.4) to (1,4.4);
\node at (0,4.8) {\textcolor{red}{$t$}};
\node at (-4.3,0) {\textcolor{red}{$a$}};
\node at (-4.3,4) {\textcolor{red}{$b$}};
\node at (0.6,0.9) {\textcolor{black}{$\mathcal{A}$}};
\node at (0.6,3.0) {\textcolor{black}{$\bar{\mathcal{A}}$}};
\draw[<-,thin,gray] (2,0) to (2,1.5);
\draw[->,thin,gray] (2,2.5) to (2,4);
\draw[-,thin,gray] (1.75,0) to (2.25,0);
\draw[-,thin,gray] (1.75,4) to (2.25,4);
\node at (2,2) {\textcolor{black}{$L$}};
\draw[<-,thin,gray] (-1,0) to (-1,0.5);
\draw[->,thin,gray] (-1,0.9) to (-1,1.5);
\draw[-,thin,gray] (-0.85,0) to (-1.15,0);
\draw[-,thin,gray] (-0.85,1.5) to (-1.15,1.5);
\node at (-1,.75) [scale=0.8]{\textcolor{black}{$\ell_\mathcal{A}$}};
\end{tikzpicture}
}
\hspace{0.2cm}
\subfloat[\textit{Conformal mapping of the strip on left to a UHP.} \label{UHP2}]
{
\begin{tikzpicture}[scale=0.5]
\draw[-,very thick,red!40] (-4,0) to (4,0);

\draw[fill=gray, draw=none, fill opacity = 0.1] (-4,0) to (4,0) to (4,4) to (-4,4);

\node at (2,2){\textcolor{darkpastelgreen}{$\cross$}};
\node at (0,-0.05) {\textcolor{black}{$\circ$}};
\node at (-4.3,0) {\textcolor{red}{$b$}};
\node at (4.3,0) {\textcolor{red}{$a$}};

\end{tikzpicture}
}
\hspace{0.2cm}
\subfloat[\textit{Conformal mapping of the UHP on left to a UHP with a wedge removed.} \label{UHP3}]
{
\begin{tikzpicture}[scale=0.5]
\draw[-,very thick,red!40] (0,0) to (4,0);
\draw[-,very thick,red!40] (-4,1) to (0,0);

\draw[fill=gray, draw=none, fill opacity = 0.1] (-4,1) to (0,0) to (4,0) to (4,4) to (-4,4);

\node at (2,2){\textcolor{darkpastelgreen}{$\cross$}};
\node at (-4.3,1) {\textcolor{red}{$b$}};
\node at (4.3,0) {\textcolor{red}{$a$}};

\end{tikzpicture}
}
}
\caption{\small{\textit{a) CFT on a strip with boundary conditions $a, b$. The time direction is horizontal, and the black dashed line is a constant time slice ($t=0$). A twist operator is inserted at the green cross, separating the subsystem $\mathcal{A}$ and its complement $\bar{\mathcal{A}}$.
b) The strip is now conformally transformed to a UHP. The black circle at the origin is a boundary condition changing operator $O_{b\rightarrow a}$. The twist operator is mapped to the green cross.
c) The configuration is finally mapped to a UHP with a wedge removed: the twisted upper half plane (TUHP). The boundary consists of two mismatched semi infinite lines, with boundary conditions $a$ and $b$. The twist operator is again mapped to the green cross.}}}
\label{2BoundaryFull}
\end{centering}
\end{figure}
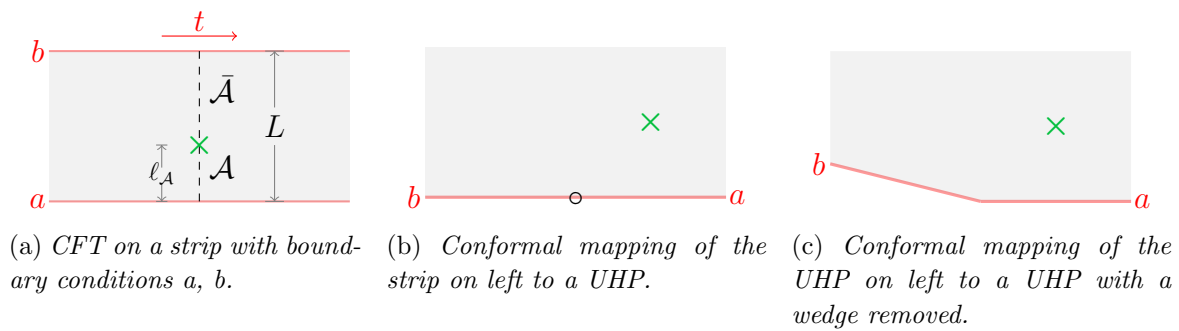

We can prepare a state of the system at a given time $t = 0$ by a Euclidean path integral from the past infinity to $t=0$.
We are interested in the entanglement entropy for a bipartition of the interval shown in Fig.~\ref{vacuum}. As 
reviewed in Sec.~\ref{sec:review}, this
reduces to computing the one-point function of the twist operator $\Phi_{n}$ inserted at the bipartition point.

Let $(w,\bar{w})$ denote complex coordinates on the strip with width $L$ so that $0<\Im(w)<L$. The subsystem $\mathcal{A}$ is of size $\ell_{\mathcal{A}} < L$.
Without loss of generality, we consider the zero Euclidean time slice, so that our inserted twist operator $\Phi_{n}(w,\bar{w})$ has $\Re(w) = \Re(\bar{w})=0$ and $\Im(w)=-\Im(\bar{w})=\ell_{\mathcal{A}}$. Now we perform a conformal transformation to map the strip with coordinates $(w,\bar{w})$ to the UHP with coordinates $(z,\bar{z})$ via
\begin{equation}
    w=\frac{L}{\pi}\ln(z) \,.
\end{equation}
The boundary $a$ is mapped to the positive real axis, and $b$ to the negative real axis, as depicted in Fig.~\ref{UHP2}.
The change in the boundary condition can be achieved by inserting a 
boundary condition changing (bcc) operator $O_{b\rightarrow a}$, which is a primary operator, at the origin, and another bcc operator $O_{a\rightarrow b}$ at $\infty$ \cite{Cardy:2004hm}.\footnote{Or equivalently, it is achieved by inserting a boundary condition changing operator $O_{b\leftarrow a}$ at the origin that changes the boundary condition on the negative real axis from $a$ to $b$ and another $O_{a\leftarrow b}$ at $\infty$.}

Inserting these bcc operators is a quench which creates energy in the system.
They can be viewed as local scaling operators with respect to the (open string) Hamiltonian $H = L_0$, performing radial evolution on the UHP \cite{Cardy:2004hm}.
Transforming to the strip, the ground state energy of this state, with respect to strip Hamiltonian, is $E = (\pi/L)(L_0 - \tfrac{c}{24}) = (\pi/L)(h_\text{bcc} - c/24)$, where $\Delta_\text{bcc} = 2h_\text{bcc}$ is the scaling dimension of the bcc operator.
%More precisely, the ground state energy of the CFT (with respect to the strip Hamiltonian) is now $E=L_{0}+\bar{L}_{0}-\frac{c}{12}=\Delta_{\text{bcc}}-\frac{c}{12}$, where $\Delta_\text{bcc}$ is the scaling dimension of the bcc operator.
%To use our knowledge of the correlators when the ground state energy is $-\frac{c}{24}$, 
We can conformally map this UHP to another UHP with conical deficit such that the energy excess $\Delta_{\text{bcc}}$ is eliminated. This is achieved by yet another conformal transformation to coordinates $(\tilde{z}, \bar{\tilde{z}})$ given by
\begin{equation}
    \tilde{z}=z^{\alpha},\quad \alpha=\sqrt{1-\frac{12\Delta_{\text{bcc}}}{c}} \,,\label{eq:Trivialization}
\end{equation}
which transforms the holomorphic energy momentum tensor to
\begin{equation}
    \widetilde{T}(\tilde{z})=\left(\frac{dz}{d\tilde{z}}\right)^2 T(z)+\frac{c}{12}\{z,\tilde{z}\}\, .
\end{equation}
The Schwarzian derivative $\{z, \tilde{z}\}$  can be computed as
\begin{equation}
    \{z,\tilde{z}\}=\frac{z'''(\tilde{z})}{z'(\tilde{z})}-\frac{3}{2}\left(\frac{z''(\tilde{z})}{z'(\tilde{z})}\right)^2=\frac{1}{2}\left(1-\frac{1}{\alpha^2}\right)\tilde{z}^{-2} \,,
\end{equation}
which gives us the holomorphic energy momentum tensor
\begin{equation}
    \widetilde{T}(\tilde{z})=\tilde{z}^{-2}\Bigg[\frac{1}{\alpha^2}z^2\, T(z)+\frac{c}{24}\left(1-\frac{1}{\alpha^2}\right)\Bigg] \,.
\end{equation}
On the $(\tilde{z},\bar{\tilde{z}})$ plane the configuration looks like Fig.~\ref{UHP3}. 
We will call this geometry the \textit{twisted upper half plane} (TUHP). We can calculate the zeroth order generator of the Virasoro algebra $\tilde{L}_0$ as
\begin{equation}
\begin{split}
    \widetilde{L}_{0}&=\int_{\mathcal{\widetilde{C}}}\frac{d\tilde{z}}{2\pi \alpha i}\tilde{z}\,\widetilde{T}(\tilde{z})\\
    &=\int_{\mathcal{C}}\frac{dz}{2\pi i}\Bigg[\frac{1}{\alpha^2}z\,T(z)+\frac{c}{24}\left(1-\frac{1}{\alpha^2}\right)\Bigg]\\
    &=\frac{1}{\alpha^2}\Big[L_{0}+\frac{c}{24}(\alpha^2-1)\Big]\: .
    \end{split}
\end{equation}
Here, in the first line, $\mathcal{\widetilde{C}}$ is the contour $\tilde{z}=e^{i\tilde{\theta}}$,  $\tilde{\theta}\in[0,2\pi\alpha]$ and $\mathcal{C}$ is the unit circle $z=e^{i\theta},\theta\in[0,2\pi]$. 
%A similar transformation happens for the anti-holomorphic energy momentum tensor.
As a result, we have the transformed Hamiltonian $\widetilde{H}$ given as
% \begin{equation}
% \begin{split}
%     \alpha^2\widetilde{H}&=\alpha^2(\widetilde{L}_0+\widetilde{\bar{L}}_0-\frac{c}{12})\\&=L_0+\bar{L}_0+\frac{c}{12}(\alpha^2-1)-\frac{c\alpha^2}{12}\\&=L_0+\bar{L}_0-\frac{c}{12}\,,
%     \end{split}
% \end{equation}
\begin{equation}
\begin{split}
    \alpha^2\widetilde{H}&=\alpha^2\frac{\pi}{L}\left(\widetilde{L}_0-\frac{c}{24}\right)\\&=\frac{\pi}{L}\left(L_0+\frac{c}{24}(\alpha^2-1)-\frac{c\alpha^2}{24}\right)\\&=\frac{\pi}{L}\left(L_0-\frac{c}{24}\right)\,,
    \end{split}
\end{equation}
which tells us that for $\widetilde{H}$ to be in its vacuum state, we need $L_{0}=\frac{c}{24}(1-\alpha^2)=\frac{\Delta_{\text{bcc}}}{2}$. This indeed corresponds to the bcc operator insertion at the origin and the infinity of the UHP.

Therefore, we see explicitly that the change of coordinates ensures that the ground state energy from the bcc operator insertion and the contribution from the conical deficit cancel. Thus, the one-point function for the twist operator $\Phi_{n}(w,\bar{w})$ becomes
\begin{equation}\begin{aligned}\label{eq:1pointTHUP}
    \langle\Phi_{n}(w,\bar{w})\rangle_{\text{strip}}^{a,b}&=\abs{\frac{dz}{dw}}^{d_{n}}\langle O_{a\leftarrow b}(\infty,\infty) \Phi_{n}(z,\bar{z})O_{b\leftarrow a}(0,0)\rangle_{\text{UHP}}^{a} \\
    &=\abs{\frac{dz}{dw}}^{d_{n}}\abs{\frac{d\tilde{z}}{dz}}^{d_{n}}\langle\Phi_{n}(\tilde{z},\bar{\tilde{z}})\rangle_{\text{THUP}}^{a,b} \\
    &=\abs{\frac{\alpha\pi}{L}e^{\frac{\alpha\pi}{L}w}}^{d_{n}}\langle\Phi_{n}(\tilde{z},\bar{\tilde{z}})\rangle_{\text{THUP}}^{a,b} \,.
\end{aligned}\end{equation}
Although the local scaling behavior of bcc operators have been trivialized from a kinematic perspective, we still have two boundary conditions.
This is because the boundary condition changing operator is not a standard local operator in CFT \cite{Cardy:2004hm} and the local physics near the boundary  
should still be controlled by the corresponding boundary conditions.
We know that on the THUP we have two intersecting boundaries, so for a bulk CFT operator, we have \textit{two} boundary channels in which to perform BOE and theses two channles should match. 

For a generic CFT, we expect that the one-point function in the presence of multiple boundaries is non-universal, since boundary operators associated with $a$ will in turn be sourced by $b$. However, for a holographic CFT the calculation simplifies due to the vacuum dominance. To be conservative,
we will treat this as a field theory ansatz for the holographic CFT on a strip, and emphasize that a full understanding of these multi-boundary correlators for a generic CFT remains an open question. This can be studied by constructing a crossing equation for the matching of the two boundary channels and looking for solutions of this crossing equation. 

For the holographic BCFT we are interested in, we compute, as in Sec.~\ref{sec:reviewBCFTEE}, the vacuum sector contributions from the two boundary channels and take the maximum %of them 
for the one-point function. As a result, we have
\begin{equation}
\begin{split}
    \langle\Phi_{n}(w,\bar{w})\rangle_{\text{strip}}
    &=\abs{\frac{\alpha\pi}{L}e^{\frac{\alpha\pi}{L}w}}^{d_{n}}\max\!\Bigg(\frac{\mathcal{B}^{a}_{\Phi_{n},1}}{\abs{2\Im\tilde{z}}^{d_{n}}},\frac{\mathcal{B}^{b}_{\Phi_{n},1}}{\abs{2\tilde{z}\sin(\alpha\pi-\arg\tilde{z})}^{d_{n}}}\Bigg)\\
    &=\abs{\frac{\alpha\pi}{L}e^{\frac{\alpha\pi}{L}w}}^{d_{n}}\max\!\Bigg(\frac{\mathcal{B}^{a}_{\Phi_{n},1}}{\abs{2e^{\frac{\alpha\pi}{L}w}\sin\frac{\alpha\pi}{L}\Im w}^{d_{n}}},\frac{\mathcal{B}^{b}_{\Phi_{n},1}}{\abs{2e^{\frac{\alpha\pi}{L}w}\sin\alpha(\pi-\frac{\pi}{L}\Im w)}^{d_{n}}}\Bigg)\\
    &=\abs{\frac{\alpha\pi}{L}}^{d_{n}}\max\!\Bigg(\frac{\mathcal{B}^{a}_{\Phi_{n},1}}{\abs{2\sin\frac{\alpha\pi}{L}\Im w}^{d_{n}}},\frac{\mathcal{B}^{b}_{\Phi_{n},1}}{\abs{2\sin\alpha(\pi-\frac{\pi}{L}\Im w)}^{d_{n}}}\Bigg) \,.
    \end{split}
\end{equation}
From Eq.~\eqref{eq:von}, the entanglement entropy is
\begin{equation}
    \begin{split}
        S_{\mathcal{A}}=\min\Bigg[\frac{c}{6}\ln(\frac{2 L}{\alpha\pi\epsilon}\sin\frac{\alpha\pi \ell_{\mathcal{A}}}{L})+\ln(g_{a}),\,\frac{c}{6}\ln(\frac{2 L}{\alpha\pi\epsilon}\sin\frac{\alpha\pi (L-\ell_{\mathcal{A}})}{L})+\ln(g_{b})\Bigg] \,, \label{eq:EEBCFTv}
    \end{split}
\end{equation}
where we used the fact that $\Im w=\ell_{\mathcal{A}}$ and the definition of boundary entropy in Eq.~\eqref{eq:bdyentropy}. So far, we have implicitly assumed that $\alpha$ is real or $\Delta_{\text{bcc}}<\frac{c}{12}$. However, generically $\alpha$ can be imaginary and this changes the 
scaling behavior of the entanglement entropy with the subsystem size $\ell_{\mathcal{A}}$ from $\sin\frac{\abs{\alpha}\pi \ell_{\mathcal{A}}}{L}$ to $\sinh\frac{\abs{\alpha}\pi \ell_{\mathcal{A}}}{L}$. This tells us that when $\Delta_{\text{bcc}}$ is large enough the field theory state is locally thermalized.\footnote{This can be seen by observing that for a CFT$_{2}$ in thermal state, the entanglement entropy scales with the subsystem size $l$ as $\ln(\frac{\beta}{\pi\epsilon}\sinh(\frac{\pi l}{\beta}))$ \cite{Ryu:2006ef,Calabrese:2009qy}. } This provides an important hint about the dual gravitational description.

Indeed, as we will see in the next section, on the gravity side the finite $L$ result will be reproduced by looking at a point particle in the global AdS$_{3}$ (the defect AdS$_{3}$ geometry) for real $\alpha$ and by considering an AdS$_{3}$ black hole for imaginary $\alpha$.

\subsection{CFT with two boundaries, at non-zero temperature}\label{sec:BCFTEE}

\begin{figure}[h]
\begin{centering}
{
\begin{tikzpicture}[scale=1.4]
\draw[-,color=red, thick] (-1,0) arc (1:361:-1);
\draw[-,color=red, thick] (-3,0) arc (1:361:-3);
\draw[-,dashed,color=black, thin] (-3,0) to (-1,0);
\draw[-,dashed,color=black, thin] (1,0) to (3,0);
\node at (-2,0.2)
{\textcolor{black}{BCFT$_L$}};
\node at (2,0.2)
{\textcolor{black}{BCFT$_R$}};

%\draw[-,dashed,color=darkpastelgreen, thick](0.705,0.705) to (1.41+0.705,1.41+0.705);
%\draw[-,dashed,color=darkpastelgreen, thick](-0.705,0.705) to (-1.41-0.705,1.41+0.705);

\draw[-,dashed,color=darkpastelgreen, thick](0.705+0.5,0.705+0.5) to (1.41+0.705,1.41+0.705);
\draw[-,dashed,color=darkpastelgreen, thick](-0.705-0.5,0.705+0.5) to (-1.41-0.705,1.41+0.705);
\draw[-,color=darkpastelgreen, thick](0.705,0.705) to (0.705+0.5,0.705+0.5) ;
\draw[-,color=darkpastelgreen, thick](-0.705,0.705) to (-0.705-0.5,0.705+0.5);

\draw[<-,, color=red, thin] (0.7,0.4) arc (25:-25:1);
\node at (0.45,0.2) {\textcolor{black}{$t_R$}};

\draw[<-,, color=red, thin] (-0.7,0.4) arc (-25:25:-1);
\node at (-0.45,0.2) {\textcolor{black}{$t_L$}};

\node at (0.705+0.5,0.705+0.5) {\textcolor{blue}{$+$}};
\node at (-0.705-0.5,0.705+0.5) {\textcolor{blue}{$+$}};
\node at (0.7,0.705+0.4) {\textcolor{black}{$\mathcal{A}_R$}};
\node at (-0.7,0.705+0.4) {\textcolor{black}{$\mathcal{A}_L$}};
\node at (1.5,2) {\textcolor{black}{$\bar{\mathcal{A}}_R$}};
\node at (-1.5,2) {\textcolor{black}{$\bar{\mathcal{A}}_L$}};
\node at (-3.2,0) {\textcolor{red}{$a$}};
\node at (-1.2,-0.2) {\textcolor{red}{$b$}};
\draw[->,color=gray, thin](0, 0) to (2,-2.2);
\draw[->,color=gray, thin](0, 0) to (-0.6,-0.75);
\draw[->,color=gray, thin] (0, -1.75) to (0, -1) ;
\draw[->,color=gray, thin] (0, -2.25) to (0, -2.95) ;
\node at (0,-2) {\textcolor{black}{$L$}};
\node at (1.85,-1.7) {\textcolor{black}{$r_O$}};
\node at (-0.1,-0.5) {\textcolor{black}{$r_I$}};
\end{tikzpicture}
}
\caption{\small{\textit{The two-dimensional CFT on a strip at a finite temperature: the two BCFTs L and R are in the TFD state. Time evolution is rotation with respect to the origin, and the red circles are the time evolution of the two boundaries. The boundary conditions are specified as $a$ and $b$. The black dashed line is the $t=0$ slice, and the choice of time evolution is indicated. We are interested in the entanglement entropy of the subsystem $\mathcal{A} = \mathcal{A}_L \cup \mathcal{A}_R$ corresponding to the solid green line segments connecting the inner boundary and the blue crosses. 
}}}
\label{TFDpre2}
\end{centering}
\end{figure}
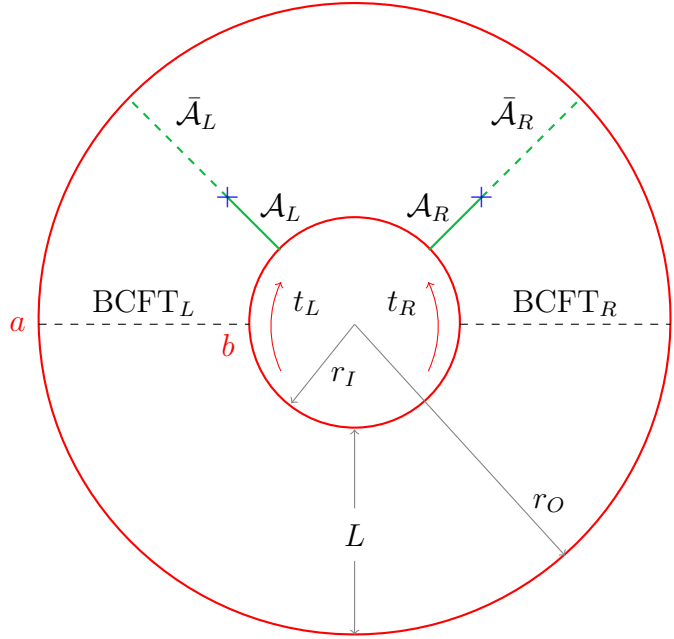

We next consider the CFT living on the strip, now at a non-zero temperature. As in subsection~\ref{sec:ReviewTFD}, we consider the thermal field double (TFD) state of two BCFTs (L and R) each with two boundaries. Both BCFTs have the same boundary conditions. This state can be easily prepared using the Euclidean path integral, after making the time direction periodic (see Fig.~\ref{TFDpre2}). This leads to considering the CFT on an annulus, with the inner and the outer circles generated by the periodic time evolution of the two boundaries of the CFT. We will denote the radii as $r_I$ and $r_O$ respectively (for inner and outer). The length of the strip is given by $L = r_O- r_I$.

To proceed, we could naively attempt for a conformal mapping to a strip and then use the technical results of the previous subsection. 
However, no conformal transformation can accomplish this, because the annulus has a modular parameter which is generally nonzero and should be preserved under any conformal transformation \cite{Polchinski:1998rq}. Consequently, there is no conformal transformation from a generic annulus to a strip. It is still possible to make progress if we focus on the high temperature limit. Specifically, we can work in the limit of fixed $L = r_O- r_I$ and 
send the radius of the smaller circle $r_I$ to be close to zero ($r_{I}/L \to 0$) in Fig.~\ref{TFDpre2}. In this limit, we \textit{can} conformally map the annulus (in $w$ coordinates) to the UHP (in $z$ coordinates) by the transformation,
\begin{equation}
    w=r_{I}\left(\frac{1}{z-i/2}-i\right) \,.\label{eq:TFDmap2}
\end{equation}
In the limit $r_I/L \rightarrow 0$, this transformation maps the inner boundary of the annulus to the real axis and its outer boundary to $z = i/2$. This is similar to Fig.~\ref{TFDmap1}, but with an operator $\Phi_{a}$ inserted at $z = i/2$, corresponding to the outer boundary of the annulus.

Now we focus on holographic CFTs which satisfy the assumptions reviewed in Sec.~\ref{sec:review}. On the UHP there are various channels for the two-point function:
\begin{equation}
    \langle\Phi_{n}(z_{R},\bar{z}_{R})\bar{\Phi}_{n}(z_{L},\bar{z}_{L})\rangle_{\text{UHP},a}^{b} \,,
\end{equation}
where the upper index $b$ denotes the boundary condition $b$ along the real axis and the lower index $a$ denotes the operator insertion $\Phi_{a}$ at $z = i/2$.
We refer to the one-point function with the shorthand $\langle\Phi_a\rangle^b = \langle\Phi_a(z_a,\bar{z}_1)\rangle_{\text{UHP}}^b$.
Here we list the channels:
\begin{itemize}
    \item The boundary channel which, up to the one-point function $\langle\Phi_a\rangle^b$, %$\langle\Phi_{a}(\frac{i}{2},-\frac{i}{2})\rangle_{\text{UHP}}^{b}$, 
    gives us 
    \begin{equation}
        \frac{\mathcal{B}_{\Phi_{n}1}^{b}\mathcal{B}_{\Phi_{n}1}^{b}}{|z_{L}-z_{L}^{*}|^{d_{n}}|z_{R}-z_{R}^{*}|^{d_{n}}} \,.
    \end{equation}
    \item The connected bulk channel which, up to the constant $\langle\Phi_a\rangle^b$, yields
    %a constant-$\langle\Phi_{a}(\frac{i}{2},-\frac{i}{2})\rangle_{\text{UHP}}^{b}$, gives us 
    \begin{equation}
        \frac{\epsilon^{2d_{n}}}{|z_{L}-z_{R}|^{2d_{n}}} \,.
    \end{equation}
    \item The disconnected bulk channel which is given by the three-point function
    \begin{equation}
        \langle\Phi_{n}(z_{L},\bar{z}_{L})\bar{\Phi}_{n}(z_{R},\bar{z}_{R})\Phi_{a}(z_a,\bar{z}_a)\rangle \,.%(\frac{i}{2},-\frac{i}{2})\rangle.
    \end{equation}
\end{itemize}

The last channel is a correlator on the whole complex plane, but can be reinterpreted as the boundary channel associated with $a$.
This can be easily computed using another conformal transformation to exchange the large and the small circle.\footnote{More precisely, we map the larger circular boundary to the real axis and the smaller one to the point $x=0,\,y=\frac{1}{2}$. This is achieved by firstly transforming the annulus to another annulus such that the two circular boundaries exchange ($w'=\frac{r_{I}r_{O}}{w}$) and then we do the same conformal transform as in Eq.~\eqref{eq:TFDmap2}.}
As reviewed in Sec.~\ref{sec:review}, for holographic CFTs we just take the maximum of these vacuum contributions to compute the entanglement entropy. Mapping back to the annulus, we get the entanglement entropy:
\begin{equation}
    \begin{split}
        S_{\mathcal{A}}=\min\Bigg[\frac{c}{3}\ln(\frac{r^2-r_{I}^2}{r_{I}\epsilon})+2\ln(g_{b}),\,\frac{c}{3}\ln(\frac{\abs{w_{R}-w_{L}}}{\epsilon}),\,\frac{c}{3}\ln(\frac{r_{O}^2-r^2}{r_{O}\epsilon})+2\ln(g_{a})\Bigg]\, .
    \end{split}
\end{equation}
Taking $w_{L}=re^{i\theta}, w_{R}=re^{i\pi-i\theta}$, and continuing back to the Lorentzian time $t$ using $t=\frac{i\theta}{2\pi}\beta$, we get
\begin{equation}
    \begin{split}
        S_{\mathcal{A}}=\min\Bigg[\frac{c}{3}\ln(\frac{r^2-r_{I}^2}{r_{I}\epsilon})+2\ln(g_{b}),\,\frac{c}{3}\ln(\frac{2r\cosh \frac{2\pi t}{\beta} }{\epsilon}),\,\frac{c}{3}\ln(\frac{r_{O}^2-r^2}{r_{O}\epsilon})+2\ln(g_{a})\Bigg] \,.\label{eq:CFTEE2}
    \end{split}
\end{equation}
This gives us a Page curve for a typical $r_{I}<r<r_{O}$. The monotonically increasing contribution from $\cosh(2\pi t/\beta)$ is the dominant contribution at early times, but is switched off at late times due to the $\min$ condition. We emphasize that this CFT calculation depends on taking the high temperature limit, the holographic assumption of vacuum dominance, and finally, the simple two-boundary ansatz for the BOE introduced in the previous section.

\section{The gravitational computation}\label{sec:gravity}
In this section, we discuss the gravitational picture, and compute the entanglement entropy, both at zero and non-zero temperature.
For the zero temperature
case, the corresponding bulk geometry depends on whether the parameter $\alpha$ is real or imaginary (or equivalently $12\Delta_\text{bcc}$ is smaller or larger than the central charge $c$). For the real case, the geometry is a point particle living on global AdS$_{3}$. For the imaginary case, the geometry is a one-sided AdS$_{3}$ black hole. In both cases we have two KR branes, separated by an arc. For the non-zero temperature case, the bulk geometry is in general complicated. However, since we are restricting to the high temperature limit, the geometry is simply a planar BTZ black hole \cite{Maldacena:2001kr,Banados:1992wn}. 

Before diving into the computation, we start with a brief review of the realization of the quantum extremal surface prescription \cite{Engelhardt:2014gca} in the KR braneworld \cite{Geng:2020qvw,Geng:2020fxl}, which is relevant for a holographic computation of the entanglement entropy.

\subsection{Quantum extremal surface in Karch-Randall braneworld}\label{sec:reviewQES}

In this subsection we give a short summary of double holography and quantum extremal surfaces in the context of KR braneworlds.
We begin with the discussion for a single KR brane, 
which will then be extended to two branes with a finite separation, the setup relevant in this paper.

\begin{figure}[h]
\begin{centering}
\begin{tabular}{c}
\subfloat[\label{pic:double_holography_1brane}]
{
\includegraphics[scale=0.5]{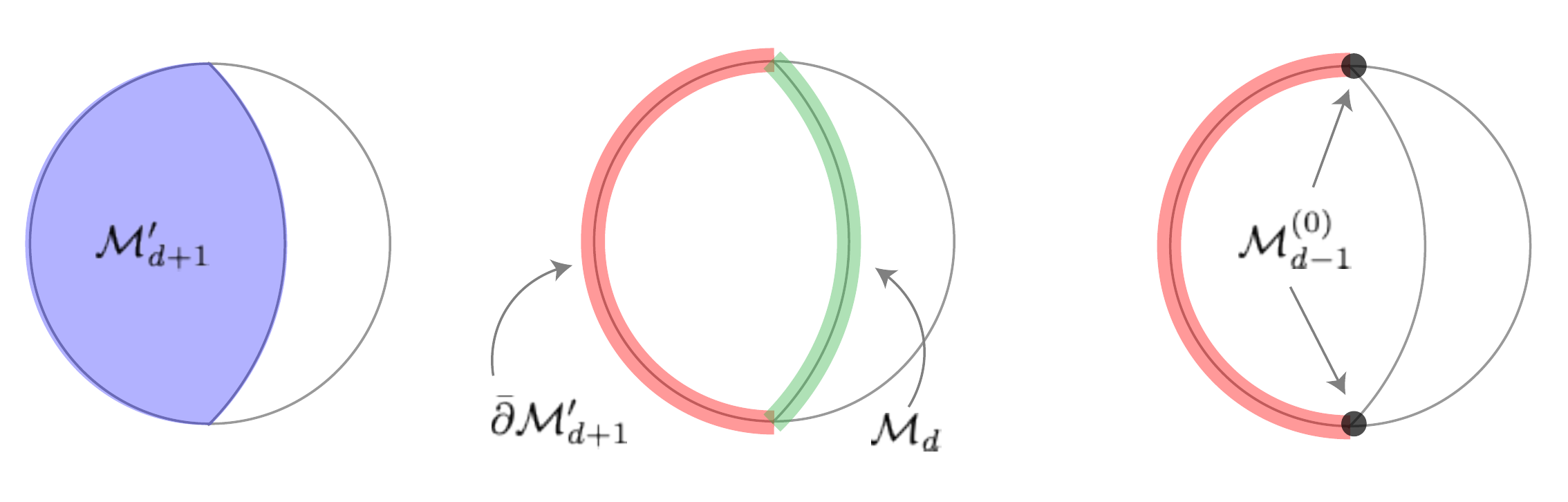}
}
\\
\subfloat[\label{pic:double_holography_2brane}]
{
\includegraphics[scale=0.5]{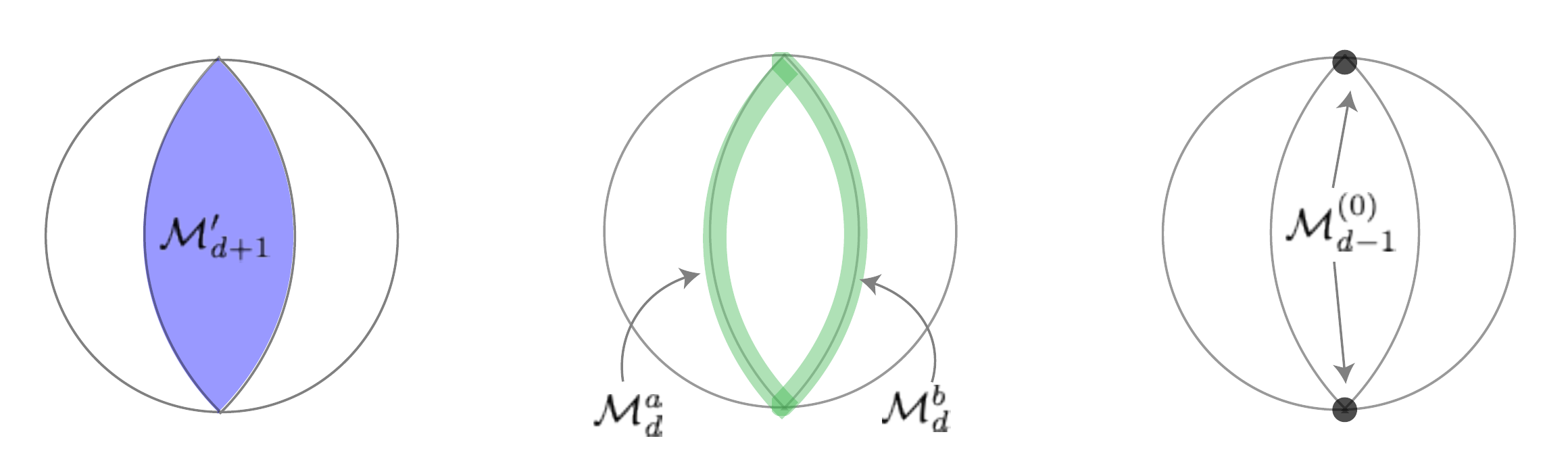}
}
\\
\subfloat[\label{pic:double_holography_2brane_bcft}]
{
\includegraphics[scale=0.5]{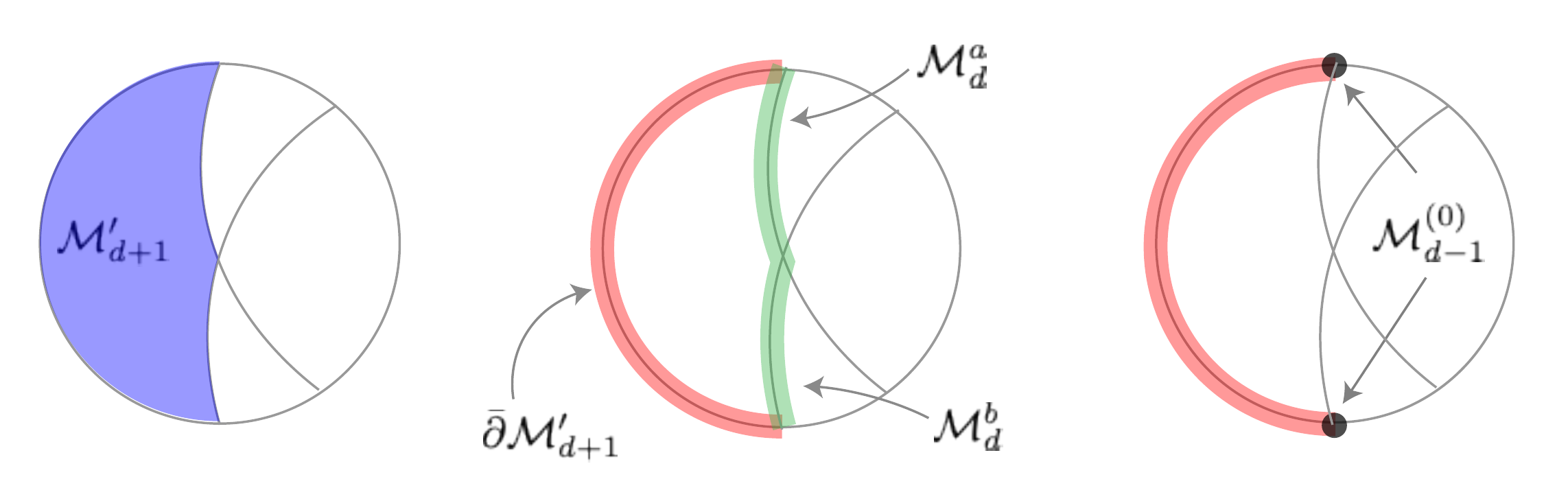}
}
\end{tabular}
\vspace{0.5cm}
\caption{\small{\textit{ Double holography in global patch with (a) one Karch-Randall brane, (b) two Karch-Randall branes with coinciding endpoints, and (c) two Karch-Randall branes with spatially separated endpoints.
The blue shaded regions $\mathcal{M}'_{d+1}$ contain a gravitational theory living on a part of an asymptotically AdS space.
The red lines represent CFTs on their asymptotic boundaries $\bar{\partial} \mathcal{M}'_{d+1}$.
The green lines $\mathcal{M}_d$, $\mathcal{M}^a_d$, and $\mathcal{M}^b_d$ are (parts of) Karch-Randall branes hosting a UV-cutoff CFT coupled to gravity.
The black dots $\mathcal{M}^{(0)}_{d-1}$ are CFTs on the asymptotic boundaries of the branes, implemented as boundary conditions of the higher-dimensional CFTs. 
In each case all three descriptions are mutually holographically dual.
In this illustration we assume that the brane tensions are positive in (a) and (b) but negative in (c). The reason for choosing negative brane tension for (c) will be made clear in the next section.
}}}
\label{pic:double_holography_full}
\end{centering}
\end{figure}

In the simplest case with a single KR brane living on an asymptotically AdS$_{d+1}$ space,\footnote{Here we keep the dimension $d$ general. In the rest of the paper, however, we exclusively consider the case $d=2$.} there exist the following three mutually dual descriptions which are depicted in Fig.~\ref{pic:double_holography_1brane}:
\begin{enumerate}
\item[(I)] \textbf{Bulk Perspective:} A $(d+1)$-dimensional gravitational theory on $\mathcal{M}'_{d+1}$, a part of an asymptotically AdS$_{d+1}$ space, containing a Karch-Randall brane $\mathcal{M}_d$ at which the bulk geometry  $\mathcal{M}'_{d+1}$ ends. 

\item[(II)] \textbf{Brane Perspective:} A CFT$_d$ living on the Karch-Randall brane $\mathcal{M}_{d}$ coupled to dynamical gravity together with a CFT$_d$ living on the asymptotic boundary $\bar{\partial}{\mathcal{M}}'_{d+1}$\footnote{Here, to avoid confusion, we use $\bar{\partial}$ to denote the asymptotic boundary and $\partial$ to denote the actual topological boundary.} of the cutoff bulk AdS$_{d+1}$. These two CFTs are connected with transparent boundary conditions on their common boundary 
$\mathcal{M}^{(0)}_{d-1} = \partial\bar{\partial}\mathcal{M}'_{d+1}=\bar{\partial} \mathcal{M}_d$.

\item[(III)] \textbf{Boundary Perspective:} A BCFT$_d$ on $\bar{\partial}{\mathcal{M}}'_{d+1}$ with a $(d-1)$-dimensional boundary $\mathcal{M}^{(0)}_{d-1}$ with conformal boundary condition imposed.
\end{enumerate}
Here, we emphasize that the geometry on the Karch-Randall brane is asymptotically AdS$_{d}$ with asymptotic boundary $\bar{\partial}{\mathcal{M}}_{d}$. These three descriptions are related by a two-fold chain of the standard AdS/CFT correspondence \cite{Maldacena:1997re,Gubser:1998bc,Witten:1998qj}, justifying the term ``double holography''.

Double holography is particularly helpful to give a holographic derivation and interpretation of the island formula \cite{Almheiri:2019hni,Geng:2020fxl}.
Consider the entanglement entropy $S(\mathcal{R})$ of a subregion $\mathcal{R}$ in the BCFT$_{d}$ in (III), disconnected from the boundary $\bar{\partial}{\mathcal{M}}_{d}$.
It can be computed using the quantum corrected Ryu-Takayanagi formula or the quantum extremal surface \cite{Lewkowycz:2013nqa,Faulkner:2013ana,Engelhardt:2014gca} in the holographically dual description (II).
This involves finding a subregion $\mathcal{I}$ within the gravitating spacetime $\mathcal{M}_{d}$ which extremizes the generalized entropy of the region $\mathcal{R}\cup\mathcal{I}$.
Moreover, if there are multiple such regions, we take the one that gives the minimum value of the generalized entropy,
\begin{equation}
    S(\mathcal{R}) = \operatorname*{min\,ext}\limits_{\mathcal{I}} \, S_\mathrm{gen} ( \mathcal{R} \cup \mathcal{I} ) \,.
\end{equation}
Since $\mathcal{I}$ is disconnected from $\mathcal{R}$, it is called an \textit{entanglement island}.
In this formula the generalized entropy functional $S_\mathrm{gen}$ is given by
\begin{equation}
    S_\mathrm{gen} ( \mathcal{R} \cup \mathcal{I} ) = \frac{A(\partial I)}{4 G_d} + S_\mathrm{mat} ( \mathcal{R} \cup \mathcal{I} ) \,,
\end{equation}
where $A(\partial I)$ denotes the area of the boundary of $\mathcal{I}$ and $G_d$ is Newton's constant of the gravitational theory on $\mathcal{M}_d$.
Moreover, $S_\mathrm{mat}$ is the entanglement entropy of matter (usually quantum field theories) in the region $\mathcal{R}\cup\mathcal{I}$.
Employing the holographic duality again, $S_\mathrm{gen}$ can be expressed in terms of a classical RT surface $\gamma$ in the bulk geometrical description (I),
\begin{equation}
    S_\mathrm{gen} ( \mathcal{R} \cup \mathcal{I} ) = \frac{A(\gamma)}{4 G_{d+1}} \,.
\end{equation}
Here, $\gamma$ is a codimension-two minimal surface in $\mathcal{M}'_{d+1}$ connecting $\partial R$ and $\partial I$ and $G_{d+1}$ is the $(d+1)$-dimensional bulk Newton's constant. As a result, we have the following formula to compute the entanglement entropy of the region $\mathcal{R}$
\begin{equation}
    S(\mathcal{R}) = \operatorname*{min\,ext}\limits_{\mathcal{I}} \,\frac{A(\gamma)}{4 G_{d+1}} \,.
\end{equation}

This setup can easily be extended to the case with a second Karch-Randall brane, a construction recently dubbed as ``wedge holography'' \cite{Akal:2020wfl,Miao:2020oey}.
The corresponding extension of the above prescription was discussed in \cite{Geng:2020fxl} for the case that the two Karch-Randall branes $\mathcal{M}^a_d$ and $\mathcal{M}^b_d$ share a common asymptotic boundary $\mathcal{M}^{(0)}_{d-1} = \bar{\partial} \mathcal{M}^a_d = \bar{\partial} \mathcal{M}^b_d$.
This implies that the $d$-dimensional BCFT in (III) degenerates to a $(d-1)$-dimensional theory living on $\mathcal{M}^{(0)}_{d-1}$.
This situation is depicted in Fig.~\ref{pic:double_holography_2brane}.

In this paper, however, we consider a slightly different configuration and allow for a finite spatial separation of the two branes at their boundaries (as illustrated in Fig.~\ref{allbcft2branes}). 
This implies that in the global coordinates the branes can overlap.
Therefore, the situation is slightly more complicated than in the previous cases, see Fig.~\ref{pic:double_holography_2brane_bcft} for details.
In particular, not only a part of the asymptotic AdS spacetime gets removed, but also the parts of the branes behind their intersection.
Moreover, the BCFT in (III) stays $d$-dimensional, but obtains another boundary: now it lives on a strip with boundaries $\bar{\partial} \mathcal{M}^a_d$ and $\bar{\partial} \mathcal{M}^b_d$.

Altogether, we arrive at the following doubly holographic description:
\begin{enumerate}
    \item[(I)] \textbf{Bulk Perspective:} A $(d+1)$-dimensional quantum gravitational theory on $\mathcal{M}'_{d+1}$, a part of an asymptotically AdS$_{d+1}$ space, containing two Karch-Randall branes.
    We denote their relevant parts in front of the intersection by $\mathcal{M}^a_d$ and $\mathcal{M}^b_d$. 
    \item[(II)] \textbf{Brane Perspective:} Two UV-cutoff CFT$_d\,$s coupled to gravity on 
    $\mathcal{M}^a_d$ and $\mathcal{M}^b_d$ together with a CFT$_d$ living on the asymptotic boundary $\bar{\partial}\mathcal{M}'_{d+1}$. These CFT$_{d}\,$s are connected with transparent boundary conditions on $\mathcal{M}^{(0)}_{d-1}$=$\bar{\partial} \mathcal{M}^a_d\cup \bar{\partial} \mathcal{M}^b_d$ (see Fig.~\ref{pic:double_holography_2brane_bcft}).
    \item[(III)] \textbf{Boundary Perspective:} A BCFT$_d$ on a $\bar{\partial}\mathcal{M}'_{d+1}$ with two $(d-1)$-dimensional boundaries $\bar{\partial} \mathcal{M}^a_d$ and $\bar{\partial} \mathcal{M}^b_d$ where conformal boundary conditions are imposed\,.
\end{enumerate}

Now we want to compute the entanglement entropy associated with a bipartition of the BCFT$_{d}$ in the description (III) such that the bipartition manifold $\partial\mathcal{R}$ is parallel to $\bar{\partial} \mathcal{M}^a_d$ and $\bar{\partial} \mathcal{M}^b_d$. This is exactly the entanglement entropy we computed in Sec.~\ref{sec:CFT} using field theory techniques. In this case, the quantum corrected Ryu-Takayanagi formula or the quantum extremal surface is double-holographically realized as
\begin{equation}
      S = \operatorname*{min\,ext}\limits_{\mathcal{I}} \,\frac{A(\gamma)}{4 G_{d+1}} \,,
\end{equation}
where $\mathcal{I}$ is a disconnected region, from the asymptotic boundary $\bar{\partial}\mathcal{M}'_{d+1}$, that lives on either of the two Karch-Randall branes and $\gamma$ is a minimal area surface in the bulk $\mathcal{M}'_{d+1}$ that connects the boundary of the island $\partial\mathcal{I}$ and the bipartition manifold $\partial\mathcal{R}$.

\subsection{Two boundaries at zero temperature}

In this subsection, we calculate the holographic entanglement entropy of a subinterval of the strip gravitationally. The system is at zero temperature.
As discussed in Sec.~\ref{sec:Vacuum}, 
we consider different 
boundary conditions on the two edges of the strip.
To determine the correct holographic dual, we note that generically the ground state energy of the strip is bigger than the vacuum value, % $(-c/12)$, 
due to the boundary conditions on the two ends being different. This difference in the boundary conditions generically results in the fields acquiring gradients between the edges, which contributes to the energy of the system.
In the CFT language, we have inserted a bcc operator so we are no longer in the vacuum state, as elaborated in Sec.~\ref{sec:CFT}.
From the state-operator correspondence, the energy of this configuration is 
\begin{equation}
    E_{\text{vac}}^{ab}=\frac{\pi}{2L}\left(\Delta_{\text{bcc}}-\frac{c}{12}\right) \,,
\end{equation}
where $\Delta_{\text{bcc}}$ is the conformal weight (or scaling dimension) of the
bcc operator $\mathcal{O}_{b\rightarrow a}$.
We set $L = \pi/2$ for convenience.

The AdS$_{3}$/CFT$_{2}$ correspondence \cite{Carlip:2005zn} tells us that the dual geometry is of the form
\begin{equation}
    ds^2=-(r^2-8GM)dt^2+\frac{dr^2}{r^2-8GM}+r^2 d\theta^2 \,, \label{eq:defectgeo}
\end{equation}
where $\theta\in[0,2\pi)$ and
\begin{equation}
M=E_{\text{vac}}^{ab}=\Delta_{\text{bcc}}-\frac{1}{8G} \,.
\end{equation}
Here we have set the curvature scale to unity and used the Brown-Henneaux central charge $c=\frac{3}{2G}$ \cite{Brown:1986nw}.

\subsubsection{Bulk Geometry as Defect AdS$_3$}

Let us first consider the case when $\Delta_{\text{bcc}}< c/12$. In this case, the mass parameter 
$M < 0$, so this bulk dual contains a point particle, giving rise to a conical defect. This geometry is conformally equivalent to a global AdS$_{3}$ geometry with a conical deficit:
\begin{equation}
\begin{split}
    ds^2&=-\left(r^2+1-\frac{12\Delta_{\text{bcc}}}{c}\right)dt^2+\left(r^2+1-\frac{12\Delta_{\text{bcc}}}{c}\right)^{-1} dr^2 + r^2 \, d\theta^2\\
    &=-(\tilde{r}^2+1)d\tilde{t}^2+\frac{d\tilde{r}^2}{\tilde{r}^2+1}+\tilde{r}^2 d\tilde{\theta}^2 \,,\label{eq:twistedglobal}
    \end{split}
\end{equation}
where we have defined the new ``twisted" variables by $\tilde{t} = \mathcal{K}t$, $\tilde{r} = r/\mathcal{K}$, and $\tilde{\theta} = \mathcal{K}\theta$, where
\begin{equation}
    \mathcal{K} = \sqrt{1 - \frac{12\Delta_{\text{bcc}}}{c}} \,.
\end{equation}
Notice that the twisted angular coordinate $\tilde{\theta}$ is in the range $[0,2\pi \mathcal{K}]$ with $\mathcal{K} < 1$.

The bulk dual also contains two Karch-Randall branes corresponding to the boundary conditions $a$ and $b$. To figure out the bulk embedding of these two branes we go to the embedding space formulation of the AdS$_{3}$ geometry,
\begin{equation}
\begin{split}
    &ds^2=-dX_{0}^2-dX_{1}^2+dX_{2}^2+dX_{3}^2 \,,\\
    &X_{0}^2+X_{1}^2-X_{2}^2-X_{3}^2=1 \,.
    \end{split}
\end{equation}
For the twisted global patch Eq.~\eqref{eq:twistedglobal}, we have
\begin{equation}
    \begin{split}
        X_{0}&=\sqrt{1+\tilde{r}^2}\cos{\tilde{t}} \,,\\
        X_{1}&=\sqrt{1+\tilde{r}^2}\sin\tilde{t} \,,\\
        X_{2}&=\tilde{r}\cos\tilde{\theta} \,,\\
        X_{3}&=\tilde{r}\sin\tilde{\theta} \,,\label{eq:AdSglobal}
    \end{split}
\end{equation}
where $\tilde{r}\in[0,\infty)$. Moreover, we notice that an AdS$_{3}$ space can be foliated by AdS$_{2}$ slices (corresponding to constant $R$),
\begin{equation}
    \begin{split}
        X_{0}&=\cosh(R)\sqrt{1+\rho^2}\cos(t) \,,\\
        X_{1}&=\cosh(R)\sqrt{1+\rho^2}\sin(t) \,,\\
        X_{2}&=\cosh(R)\rho \,,\\
        X_{3}&=\sinh(R) \,,
    \end{split}\label{eq:AdS/adS}
\end{equation}
for $\rho\in(-\infty,\infty)$.
Comparing these two parametrizations we find
\begin{equation}
    \tan\tilde{\theta}=\frac{\tanh(R)}{\rho} \,,
\end{equation}
which tells us that each AdS$_{2}$ slice is stretched between two boundary points with an angular separation $\Delta\tilde{\theta}=\pi$. In terms of the untwisted coordinates this implies $\Delta\theta = \pi/\mathcal{K}$, 
which is larger than $\pi$ for $c/12>\Delta_{\text{bcc}}\neq0$.

\begin{figure}
\begin{centering}
{
\begin{tikzpicture}[scale=0.7]
\draw[-,color=blue,very thick] (3,0) arc (0:361:3);
\draw[-,color=red,thick] (0,3) arc (120:283:2.8);
\draw[-,color=darkpastelgreen,thick] (0,-3) arc (231:92:3.02);
\draw[-,color=gray,thick] (-1.5,2.61) arc (225:242:2.5);
\draw[-,color=gray,thick,dashed] (-1.5,2.61) arc (200:232:3);
\node at (-1.5,2.61) {\textbf{\textcolor{black}{$+$}}};
\node at (0,3) {\textcolor{red}{$\bullet$}};
\node at (0,-3) {\textcolor{darkpastelgreen}{$\bullet$}};
\node at (0,3.5) {\textcolor{red}{$a$}};
\node at (0,-3.5) {\textcolor{darkpastelgreen}{$b$}};
\draw[pattern=north west lines,pattern color=gray,draw=none] (0,3) arc (120:283:2.8)--(0,3) arc (90:-45:3);
\draw[pattern=north west lines,pattern color=gray,draw=none] (0,-3) arc (231:92:3.02)--(0,-3) arc (-90:52:3);
\end{tikzpicture}
}
\caption{\small{\textit{A constant time slice of the point particle (defect) bulk geometry Eq.~\eqref{eq:defectgeo} in the untwisted radial and angular coordinates $(r,\theta)$ with two Karch-Randall branes. The blue circle is the bulk conformal boundary.
The two conformal boundaries $a$ and $b$ of the branes (in red and green) are separated by $\pi$ in the coordinate $\theta$ ($\theta_{a}=0$ and $\theta_{b}=\pi$). The black cross is an example bipartition point on the boundary with two Ryu-Takayanagi surfaces ending on the branes. The physical RT surface (solid) has always a smaller area than the unphysical one (dashed) ending on the ``hidden'' part of the 
bulk.}}}
\label{pic:defectbrane}
\end{centering}
\end{figure}
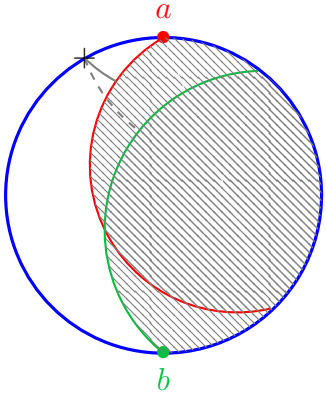

The two branes correspond to two such AdS$_2$ slices, with the following embeddings:
\begin{equation}
\begin{split}\label{eq:branes}
    \text{Brane } a\colon& \tan\tilde{\theta}=\frac{\tanh(R_{a})}{\rho} \,,\\
    \text{Brane } b\colon& \tan(\widetilde{\theta-\pi}+\pi)=\frac{\tanh(R_{b})}{\rho} \,,
\end{split}
\end{equation}
where 
$\widetilde{\theta-\pi} =\tilde\theta-\pi\mathcal{K}$.
The two AdS$_2$ foliations in which brane $a$ and brane $b$ are constant $R$ slices are oppositely oriented and not aligned.
Therefore, we need to set $R_{a}>0$ and $R_{b}<0$.
For definiteness, we restrict the domain of the $\tan$ function to $[0,\pi] \backslash \{\frac{\pi}{2}\}$.
This implies that brane $a$ stretches from $\theta = 0$ to $\theta=\frac{\pi}{\mathcal{K}}$ and brane $b$ stretches from $\theta =\pi$ to $ \theta=\pi-\frac{\pi}{\mathcal{K}}$.%
\footnote{ Notice, that using this embedding, parts of brane $b$ are at negative values of $\tilde \theta$.
Exploiting the $2 \pi \mathcal{K}$ period of $\tilde \theta$, this segment can be mapped into the previously specified domain $\tilde \theta \in [0, 2 \pi \mathcal{K} ]$.
Strictly speaking however, 
it is better to describe the brane $b$ using a different fundamental domain for $\tilde{\theta}$ and at the end exploiting the $2\pi\mathcal{K}$ period to match with the $[0,2\pi\mathcal{K}]$ domain.}
This configuration is illustrated in Fig.~\ref{pic:defectbrane}.
To avoid self-intersection for each brane, we require that $\mathcal{K}\geq\frac{1}{2}$ which translates to
\begin{equation}
    \Delta_{\text{bcc}}\leq \frac{c}{16} \,.\label{eq:bound1}
\end{equation}
We conclude that there is a gap $(\frac{c}{16},\frac{c}{12})$ in the spectrum of $\Delta_{\text{bcc}}$.

 Such branes satisfy Einstein's equations if they have a specific tension $T$ which can be determined from the junction condition \cite{Kraus:1999it} 
\begin{equation}\label{eq:junction}
    K_{\mu\nu} = T h_{\mu\nu} \,,
\end{equation}
where $h_{\mu\nu}$ is the induced metric on the brane and $K_{\mu\nu}$ is its extrinsic curvature \cite{Wald:1984rg}.
The latter can be expressed as the derivative of the induced metric with respect to the brane's unit normal vector $n^\mu$,
\begin{equation}\label{eq:extrinsic}
K_{\mu\nu} = h_{\mu}^{\rho}h_{\nu}^{\sigma}\nabla_{\rho}n_{\sigma} \,.
\end{equation}
As explained above, we consider a situation where the bulk spacetime ends at the branes and the branes are located such that the conformal boundary of the bulk stretches from $\theta = 0$ to $\theta = \pi$, i.e.~from the first endpoint of brane $a$ to the opposite-lying endpoint of brane $b$ (see Fig.~\ref{pic:defectbrane}). 
To satisfy \eqref{eq:junction} the normal vector $n^\mu$ has to be chosen in such a way that it points in the outward direction from the point of view of the bulk spacetime ending on the branes.
We can then compute $K_{\mu\nu}$ and $h_{\mu\nu}$ explicitly and find
\begin{equation}
    T_{a} = -\tanh(R_{a}) \qquad\text{and}\qquad  T_{b}=\tanh(R_{b}) \,.
\end{equation}
The relative minus sign comes from the fact that the two AdS$_2$ foliations in which brane $a$ and brane $b$ are described are oppositely oriented. 
For the same reason, the brane embeddings Eq.~\eqref{eq:branes} require $R_{a} > 0$ and $R_{b}<0$, so that both branes have negative tension.

To study the RT surface, we use again the AdS$_{2}$ foliation Eq.~\eqref{eq:AdS/adS}. In these coordinates, the bulk metric reads
\begin{equation}
    ds^2=dR^2+\cosh^{2}(R)\left(-(\rho^2+1)dt^2+\frac{d\rho^2}{\rho^2+1}\right) \,,
\end{equation}
and the two Karch-Randall branes are two constant $R$ slices. 
We let $\rho=\sinh(\eta)$ and parametrize the RT surface by $\eta(R)$. The area functional for the RT surface ending on $a$ then reads
\begin{equation}
    A_{a}=\int_{R_{a}}^{R_{\epsilon}}dR \sqrt{1+\cosh^{2}(R)\eta'(R)^2} \,,
\end{equation}
where $R_{\epsilon}$ is a UV regulator. The resulting Euler-Lagrangian equation can be integrated to 
\begin{equation}
    C=\frac{\cosh^{2}(R)\eta'(R)}{1+\cosh^{2}(R)\eta'(R)^2} \,,
\end{equation}
for a constant $C$ determined by the boundary condition for the RT surface near its end point on the brane $a$. This boundary condition, worked out in \cite{Geng:2020fxl}, is
\begin{equation}
    \left|\eta'(R)\right|_{\text{brane } a}=0 \,.
\end{equation}
Therefore, we have $C=0$ and the area
\begin{equation}
    A_{a}=R_{\epsilon}-R_{a} \,.
\end{equation}
This is an attempt to calculate the entanglement entropy of the boundary subsystem $\mathcal{A}$ with size $\theta$. The UV regularization $R_{\epsilon}$ depends on the size of this boundary subsystem, and can be extracted by comparing Eqs.~\eqref{eq:AdSglobal} and \eqref{eq:AdS/adS},
\begin{equation}
    1+\tilde{r}^2\sin^{2}(\tilde{\theta})=\cosh^{2}(R) \,.
\end{equation}
Hence, we have
\begin{equation}
    \frac{4}{\epsilon^2} \sin^{2}(\tilde{\theta})=e^{2R_{\epsilon}} \,,
\end{equation}
where we take $\tilde{r}=\epsilon^{-1}\rightarrow\infty$ and $R
_\epsilon \rightarrow \infty$. Using the RT formula \cite{Ryu:2006bv} 
and the Brown-Henneaux central charge \cite{Brown:1986nw} we get a candidate 
expression for the entanglement entropy of the boundary subsystem $\mathcal{A}$,
\begin{equation}
    S_{\mathcal{A}}^{a}=\frac{A_{a}}{4G}=\frac{c}{6}\ln\left(\frac{2}{\epsilon}\sin(\theta\mathcal{K})\right)%\ln(2\frac{\sin(\theta\sqrt{1-\frac{24\Delta_{\text{bcc}}}{c}})}{\epsilon})
    +\ln(g_{a}) \,,
\end{equation}
where we used the fact that $-R_{a}/4G=\ln(g_{a})$.
This RT surface corresponds to the solid gray line in Fig.~\ref{pic:defectbrane}.\footnote{Some care is required here. For negative-tension branes, we cut off the right part of the conformal boundary in Fig.~\ref{pic:defectbrane}, and we have $R_a>0$ and $R_b < 0$.}
Similarly, if we look at the RT surface which ends on the brane $b$ we get
\begin{equation}
    A_{b}=R_{b}-R'_{\epsilon} \,,
\end{equation}
and we have
\begin{equation}
\frac{4}{\epsilon^2}\sin^{2}(\widetilde{\theta-\pi}+\pi)=e^{-2R'_{\epsilon}} \,.
\end{equation}
Using $\frac{R_b}{4G}=\ln(g_b)$ we get another candidate expression for the entanglement entropy of the boundary subsystem $\mathcal{A}$,
\begin{equation}
    S_{\mathcal{A}}^{b}=\frac{c}{6}\ln\left(\frac{2}{\epsilon}\sin[(\pi-\theta)\mathcal{K}]\right)+\ln(g_{b}) \,.
\end{equation}
This RT surface corresponds to the dashed gray line in Fig.~\ref{pic:defectbrane}.

In summary, for $\Delta_{\text{bcc}}<c/12$, we obtain the entanglement entropy
\begin{equation}
\begin{split}
    S_{\mathcal{A}} = \min\Biggl[\frac{c}{6}\ln\left(\frac{2}{\epsilon}\sin(\theta\mathcal{K})\right) + \ln (g_a),\, \frac{c}{6}\ln\left(\frac{2}{\epsilon}\sin[(\pi-\theta)\mathcal{K}]\right) + \ln (g_b)\Biggr].
    \end{split}
\end{equation}
This exactly matches the result of our field theory calculation Eq.~\eqref{eq:EEBCFTv}, with the identifications
$\pi \ell_{\mathcal{A}}/L=\theta$ and $\mathcal{K}=\alpha$.  
We have previously established that both branes have negative tensions.
Since the brane tension determines the boundary entropy, this implies that both $\ln g_a$ and $\ln g_b$ are also negative. Furthermore, it is important to notice that the minimization prescription tells us that the actual RT surface calculating the entanglement entropy never probes into the removed region. This is provides a nontrivial consistency check of our setup (as depicted in Fig.~\ref{pic:defectbrane}).

\subsubsection{Bulk Geometry as a Single-sided Black Hole}

For $\Delta_{\text{bcc}}>c/12$ one sees from the metric Eq.~\eqref{eq:defectgeo} that the bulk dual is a single-sided black hole with mass $m=M+1/8G=\Delta_{\text{bcc}}$ and zero spin. The entanglement entropy calculation is in parallel with the defect case and the result can be simply obtained by analytic continuation of the parameter $\mathcal{K} =\sqrt{1-12\Delta_{\text{bcc}}/c}$ 
from real to imaginary. Consequently, one just has to change $\sin\to\sinh$, as on the field theory side. Hence, for $\Delta_{\text{bcc}}> c/12$ we find the entanglement entropy
\begin{equation}
\begin{split}
S_{\mathcal{A}} = \min\Biggl[\frac{c}{6}\ln\left(\frac{2}{\epsilon}\sinh(\theta\abs{\mathcal{K}})\right) + \ln (g_a),\, \frac{c}{6}\ln\left(\frac{2}{\epsilon}\sinh[(\pi-\theta)\abs{\mathcal{K}}]\right) + \ln (g_b)\Biggr] \,,\label{eq:ssBHentropy}
    \end{split}
\end{equation}
which also exactly matches the field theory result, given the fact that $\mathcal{K}=\alpha$.

\subsection{Thermal field double state}
In this subsection, we perform the holographic computation for the entanglement entropy of the subsystem $\mathcal{A}$ at a finite temperature as considered in Sec.~\ref{sec:BCFTEE}. This is very similar to the left-right entanglement entropy $S_{L/R}$ that was studied in \cite{Geng:2020fxl}, and characterizes the communication between two lower-dimensional black holes.  The two black holes are living on each of the branes, and are induced from the bulk BTZ black hole:
\begin{equation}
 ds^2=-\frac{h(z)}{z^2}dt^2+\frac{dz^2}{h(z)z^2}+\frac{dx^2}{z^2} \,,\qquad h(z)=1-\frac{z^2}{z_{h}^2} \,.  
\end{equation}
To match the field theory computation, we focus on the high temperature phase where the spatial direction $x$ is fully decompactified. We consider the maximally extended black hole spacetime with two asymptotic (conformal) boundaries, and the field theory duals 
are in the TFD state \cite{Maldacena:2001kr}. For simplicity, when we consider the branes and the RT surfaces ending on the branes, we only consider a single side of the two-sided bulk geometry.

The two KR branes are both parametrized by $x(z)$ and are chosen to have a separation $\Delta x=\ln(r_{O}/r_{I})$ on the conformal boundary $z=0$. For such a brane $x(z)$, 
the outward normal vector is
\begin{equation}
    n_{\mu}=\frac{(0,-x'(z),1)}{\sqrt{x'^2(z) h(z)z^2+z^2}} \,.
\end{equation}
As before, the shape of the brane is determined by the boundary equations of motion in the form of Israel's junction condition Eq.~\eqref{eq:junction}.
Using Eq.~\eqref{eq:extrinsic} we find
\begin{equation}
    x'(z)=\pm\frac{T}{\sqrt{1-T^2h(z)}} \,.
\end{equation}
This differential equation integrates to
\begin{equation}
    x(z)=x(0)\pm z_h \arcsinh{\left(\frac{z\,T}{z_{h}\sqrt{1-T^2}}\right)} \,.
\end{equation}
Notice that this requires that $T^2<1$, consistent with the fact that KR branes are subcritical \cite{Karch:2020iit}.

\begin{figure}[h]
\begin{centering}
\begin{tikzpicture}[scale=1.5]
\draw[-,very thick,blue!40] (-3,0) to (3,0);
\node at (-1,0) {\textcolor{red}{$\bullet$}};
\node at (-1,0.2) {\textcolor{red}{$a$}};
\node at (1,0) {\textcolor{darkpastelgreen}{$\bullet$}};
\node at (1,0.2) {\textcolor{darkpastelgreen}{$b$}};
\draw[-,dashed,very thick,black!40] (-3,-2) to (3,-2);
\draw[-,very thick,red] (-1,0) arc (100:175:2.2); 
\draw[-,very thick,darkpastelgreen] (1,0) arc (50:-9:2.2); 
\draw[fill=gray, draw=none, fill opacity = 0.1] (-3,0)--(-1,0) arc (100:175:2.2)--(-3,-2) ;
\draw[fill=gray, draw=none, fill opacity = 0.1] (3,0)--(1,0) arc (50:-9:2.2)--(3,-2) ;
\end{tikzpicture}
\caption{\small{\textit{A typical brane configuration on a constant time $t=t_{0}$ slice in the presence of a bulk BTZ black hole.
Both branes have positive tension and are embedded according to Eq.~(\ref{eq:L/Rbranes}).
The shaded regions are removed and the dashed black line is the planar black hole horizon. We neglect the parts of the branes behind the black hole horizon which are easily seen to be timelike.}}}
\label{pic:eternalbran}
\end{centering}
\end{figure}
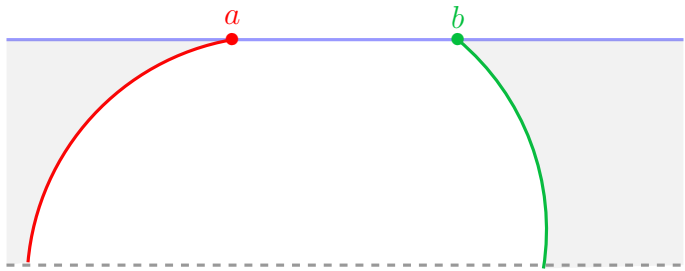

For our purposes, we take both branes to have positive tension (not to be confused with the defect case, where the tensions were negative), but one takes the positive sign and the other one takes the negative sign in the above solution,
\begin{equation}
\begin{aligned}
    x_{L}(z) &= z_{h}\ln(\frac{r_{O}}{z_{h}})+z_h \arcsinh{\left(\frac{zT_{L}}{z_{h}\sqrt{1-T_{L}^2}}\right)} \,, \\
    x_{R}(z) & =z_{h}\ln(\frac{r_{I}}{z_{h}})- z_h \arcsinh{\left(\frac{zT_{R}}{z_{h}\sqrt{1-T_{R}^2}}\right)} \,.\label{eq:L/Rbranes}
\end{aligned}
\end{equation}
A typical brane configuration with such an embedding is illustrated in Fig.~\ref{pic:eternalbran}.

We first study the RT surfaces that start from the conformal boundary and end on one of the branes. As in \cite{Geng:2020fxl}, this RT surface is time-independent, therefore we can compute it on the zero time slice with the metric
\begin{equation}
    ds^2|_{t=0}=\frac{dz^2}{h(z)z^2}+\frac{dx^2}{z^2} \,.
\end{equation}
To simplify the computation for these brane-ending RT surfaces, we perform the coordinate transformation
\begin{equation}
\tilde{x}=z_h e^\frac{x}{z_h}\sqrt{h(z)}\,,\qquad \tilde{z}=ze^{\frac{x}{z_{h}}} \,.\label{eq:transform}
\end{equation}
This maps the zero time slice onto a constant time slice in the Poincare patch,
\begin{equation}
    ds^2|_{t=0}=\frac{d\tilde{x}^2+d\tilde{z}^2}{\tilde{z}^2} \,.
\end{equation}
The branes become circles on this constant time slice, obeying
\begin{equation}
\begin{split}
    \tilde{x}_{L}^2+\left(\tilde{z}-\frac{r_{O}T_{L}}{\sqrt{1-T_{L}^2}}\right)^2=\frac{r_{O}^2}{1-T_{L}^2} \,,\\
    \tilde{x}_{R}^2+\left(\tilde{z}+\frac{r_{I}T_{R}}{\sqrt{1-T_{R}^2}}\right)^2=\frac{r_{I}^2}{1-T_{R}^2} \,.
    \end{split}
\end{equation}
We notice from Eq.~\eqref{eq:transform} that we require $\tilde{x}>0$, so strictly speaking the branes are only arcs on these circles. 

We take the bipartition point on the conformal boundary to be at $x=z_{h}\ln(r/z_{h})$, or equivalently $\tilde{x}=r$. We parametrize the RT surface ending on the left brane as $\tilde{x}(z)$ to obtain the area functional
\begin{equation}
    A=\int_{0}^{\tilde{z}_{L}}\frac{d\tilde{z}}{\tilde{z}} \sqrt{\tilde{x}'(\tilde{z})^2+1} \,,
\end{equation}
where $\tilde{z}_{L}$ is the $z$-coordinate of the end point on the brane. This leads to an Euler-Lagrangian's equation which can be integrated to
\begin{equation}
    \frac{1}{\tilde{z}_{*}}=\frac{1}{\tilde{z}}\frac{\tilde{x}'}{\sqrt{\tilde{x}'^2+1}} \,,
\end{equation}
where $\tilde{z}_{*}$ is the bulk point at which $\tilde{x}'$ goes from negative to positive. Integrating the equation of motion with the boundary condition $\tilde{x}(0)=r$ we find
\begin{equation}
    (\tilde{x}-r-\tilde{z}_{*})^2+\tilde{z}^2=\tilde{z}_{*}^2 \,,
\end{equation}
which is again a circle. 

To pin down the values of $\tilde{z}_{*}$ and $\tilde{z}_{L}$, we need the boundary condition of the RT surface on the left brane. This boundary condition is determined by area minimization over all possible end points on the brane. As suggested in \cite{Geng:2020fxl}, this is most straightforwardly determined by parametrizing the RT surface using its (normalized) proper length $s$ ($0<s<1$). Now the area functional reads
\begin{equation}
    A=\int_{0}^{1}ds\frac{1}{\tilde{z}}\sqrt{\dot{\tilde{z}}^2+\dot{\tilde{x}}^2} \,.
\end{equation}
Assuming that the Euler-Lagrange equation is satisfied, the variation of the action is
\begin{equation}
    \delta A=\left.\frac{\dot{\tilde{z}}\delta{\tilde{z}}+\dot{\tilde{x}}\delta{\tilde{x}}}{\tilde{z}\sqrt{\dot{\tilde{z}}^2+\dot{\tilde{x}}^2}}\right|_{0}^{1} \,.
\end{equation}
Demanding that this vanishes, we find that the end-point of the RT surface for the left brane obeys
\begin{equation}
    \tilde{x}'(\tilde{z}_{L})=\frac{\dot{\tilde{x}}}{\dot{\tilde{z}}}=-\frac{\delta{\tilde{z}}}{\delta{\tilde{x}}}=-\left.\tilde{z}'(\tilde{x})\right|_{\text{left brane}} \,,
\end{equation}
where $\tilde{x}'(\tilde{z}_{L})$ is the derivative of the RT surface at its end-point. Thus, we can finally determine our parameters:
\begin{equation}
    \tilde{z}_{*}=\frac{r_{O}^2-r^2}{2r} \,,\qquad \tilde{z}_{L}=\frac{2 r_{O}^{3}(1-T_L)T_L}{\sqrt{1-T_L^2}(r_{O}^2 (1-T_L)+r^2 (1+T_L))} \,.
\end{equation}
The regularized area is therefore
\begin{equation}
\begin{split}
    A_{L}&=\int_{\epsilon}^{\tilde{z}_*}\frac{d\tilde{z}\,\tilde{z}_*}{\tilde{z}\sqrt{\tilde{z}_{*}^2-\tilde{z}^2}}+\int_{\tilde{z}_{L}}^{\tilde{z}_*}\frac{d\tilde{z}\,\tilde{z}_*}{\tilde{z}\sqrt{\tilde{z}_{*}^2-\tilde{z}^2}}\\
    &=\ln(\frac{r_{O}^2-r^2}{r_{O}\epsilon}\sqrt{\frac{1+T_{L}}{1-T_{L}}}) \,.
    \end{split}
\end{equation}
The integrals are easiest to perform using the polar-parametrized equations for circles.
Similar calculations can be done for the right brane, and the area of the corresponding brane-ending RT surface is
\begin{equation}
    A_{R}=\ln(\frac{r^2-r_{I}^2}{r_{I}\epsilon}\sqrt{\frac{1+T_{R}}{1-T_{R}}}) \,.
\end{equation}
The brane tensions are related to the boundary entropies $\ln(g)$ as follows \cite{Takayanagi:2011zk}:
\begin{equation}
    \ln(g)=\frac{c}{6}\arctanh T=\frac{c}{6}\ln(\sqrt{\frac{1+T}{1-T}}).
\end{equation}
Thus, we can simplify the area results to
\begin{equation}
    A_L = \ln\left(\frac{r_{O}^2-r^2}{r_{O}\epsilon}\right) + \frac{6}{c}\ln(g_L) \,, \qquad 
    A_R = \ln\left(\frac{r^2-r_I^2}{r_{I}\epsilon}\right) + \frac{6}{c}\ln(g_R) \,. \label{eq:areas}
\end{equation}
To get the entanglement entropy for our two-sided bipartition, we double the areas of the brane ending RT surfaces. 

Hence, using the RT formula, $c=3/2G$, and Eq.~\eqref{eq:areas}, the candidate contributions to the entanglement entropy from these quantum extremal surfaces read
\begin{equation}
\begin{split}
    S_{1}&=\min\Biggl[\frac{2A_{L}}{4G},\,\frac{2A_{R}}{4G}\Biggr] 
    =\min\Biggl[\frac{c}{3}\ln(\frac{r_{O}^2-r^2}{r_{O}\epsilon})+2\ln(g_{L}),\,\frac{c}{3}\ln(\frac{r^2-r_{I}^2}{r_{I}\epsilon})+2\ln(g_{R})\Biggr].
    \end{split}
\end{equation}
However, there is another candidate entangling surface that stretches between the bipartition points on the two asymptotic boundaries. This is called the Hartman-Maldacena surface \cite{Hartman:2013qma} and it plays a central rule in the recent computations of Page curves for black holes in the KR braneworld \cite{Almheiri:2019psy,Geng:2020qvw,Geng:2020fxl}.\footnote{
This program has recently been extended into de Sitter holography. In \cite{Geng:2021wcq} a page curve for de Sitter space was obtained along similar lines.}

In our case, the area of this surface is most easily obtained using embedding space coordinates. Suppose that the bipartition point on one asymptotic boundary has coordinates
\begin{equation}
    \tilde{z}=ze^{\frac{x}{z_{h}}}=\epsilon \,, \quad \tilde{x}=z_{h}e^{\frac{x}{z_{h}}}\sqrt{h(z)}=r \,.
\end{equation}
Then its embedding space coordinates read
\begin{equation}
    \begin{split}
        X_{0}&=\sqrt{\frac{z_{h}^2}{z^2}-1}\sinh(\frac{2\pi t}{\beta}) \,, \\
        X_{1}&=\frac{z_{h}}{z}\cosh(\frac{2\pi x}{\beta}) \,, \\
        X_{2}&=\sqrt{\frac{z_{h}^2}{z^2}-1}\cosh(\frac{2\pi t}{\beta})\,, \\
        X_{3}&=\frac{z_{h}}{z}\sinh(\frac{2\pi x}{\beta}) \,,
    \end{split}
\end{equation}
where the inverse temperature of the black hole is $\beta=2\pi z_{h}$. The embedding space coordinate for the bipartition point on the other asymptotic boundary is obtained by sending $t$ to $-t+\frac{i\beta}{2}$,\footnote{This can be seen by going to the Euclidean signature, rotating around the thermal circle, and then returning to Lorentzian signature.}
\begin{equation}
    \begin{split}
        X_{0}'&=\sqrt{\frac{z_{h}^2}{z^2}-1}\sinh(\frac{2\pi t}{\beta}) \,,\\
        X_{1}'&=\frac{z_{h}}{z}\cosh(\frac{2\pi x}{\beta}) \,,\\
        X_{2}'&=-\sqrt{\frac{z_{h}^2}{z^2}-1}\cosh(\frac{2\pi t}{\beta}) \,,\\
        X_{3}'&=\frac{z_{h}}{z}\sinh(\frac{2\pi x}{\beta}) \,.
    \end{split}
\end{equation}
As a result, we get the area of the Hartman-Maldacena (HM) surface as the geodesic distance between these two points on the two asymptotic boundaries:
\begin{equation}
\begin{split}
    A_\text{HM}&=\arccosh(X_{0}X_{0}'+X_{1}X_{1}'-X_{2}X_{2}'-X_{3}X_{3}')\\&=\arccosh\left(\frac{2z_{h}^2}{z^{2}}\cosh^{2}\left(\frac{2\pi t}{\beta}\right)\right)\\&=\arccosh\left(\frac{2r^2}{\epsilon^2}\cosh^2\left(\frac{2\pi t}{\beta}\right)\right)\\&=2\ln\left(\frac{2r}{\epsilon}\cosh\left(\frac{2\pi t}{\beta}\right)\right),
    \end{split}
\end{equation}
where in the last step we used the fact that we are taking the limit $\epsilon\rightarrow0$. Therefore, using the RT formula and the Brown-Henneaux central charge, its contribution to the entanglement entropy is
\begin{equation}
    S_\text{HM}=\frac{A_\text{HM}}{4G}=\frac{c}{3}\ln(\frac{2r}{\epsilon}\cosh\left(\frac{2\pi t}{\beta}\right)) \,.
\end{equation}
Putting all our candidate minimal surfaces together, we find an expression for the holographic entanglement entropy capturing the communication between two braneworld black holes:
\begin{equation}
    S=\min\Biggl[\frac{c}{3}\ln(\frac{r_{O}^2-r^2}{r_{O}\epsilon})+2\ln(g_{L}),\,\frac{c}{3}\ln(\frac{2r}{\epsilon}\cosh\left(\frac{2\pi t}{\beta}\right)),\,\frac{c}{3}\ln(\frac{r^2-r_{I}^2}{r_{I}\epsilon})+2\ln(g_{R})\Biggr].
    \label{eq:entropy-bh-gravity}
\end{equation}
This precisely matches the result of our field theory calculation, Eq.~\eqref{eq:CFTEE2}. 

\begin{figure}
\begin{centering}
\subfloat[\label{pic:entropy-time}]
{
\includegraphics[scale=0.55]{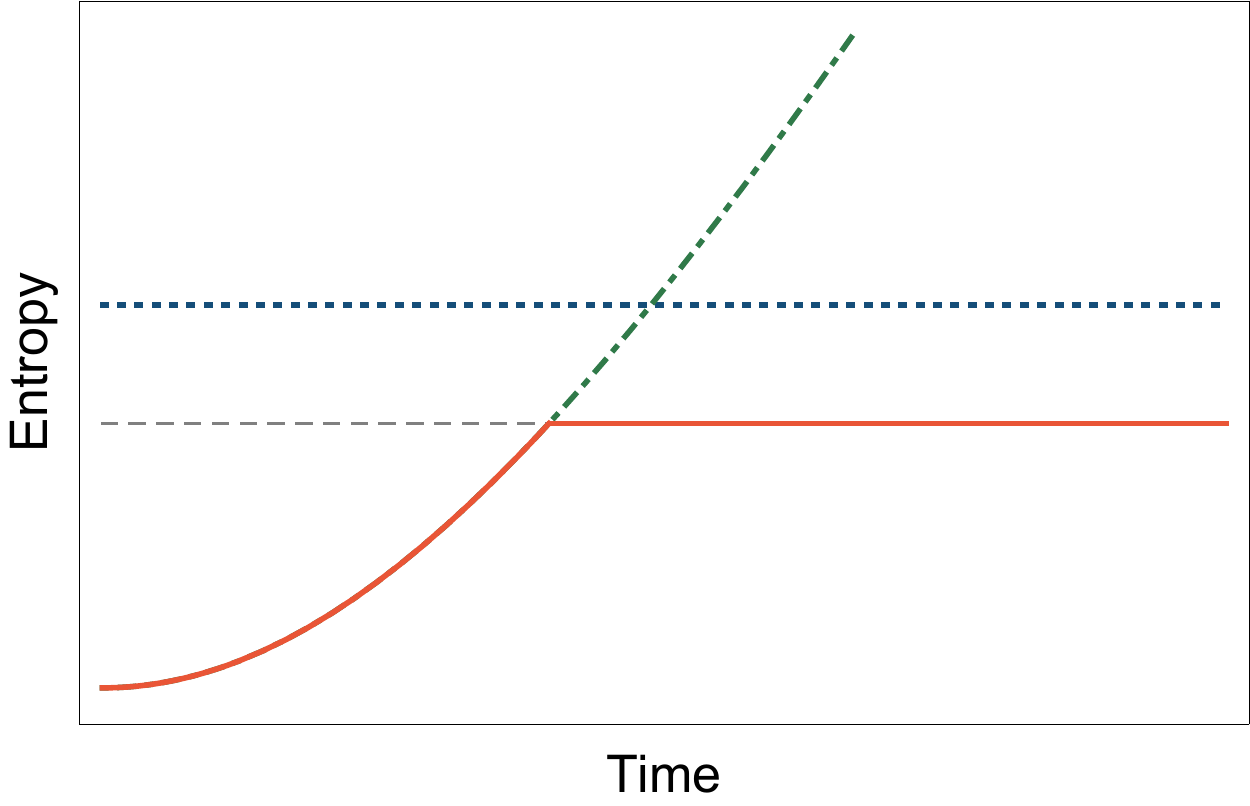}
}
\hspace{0.2 cm}
\subfloat[\label{pic:entropy-radius}]
{
\includegraphics[scale=0.55]{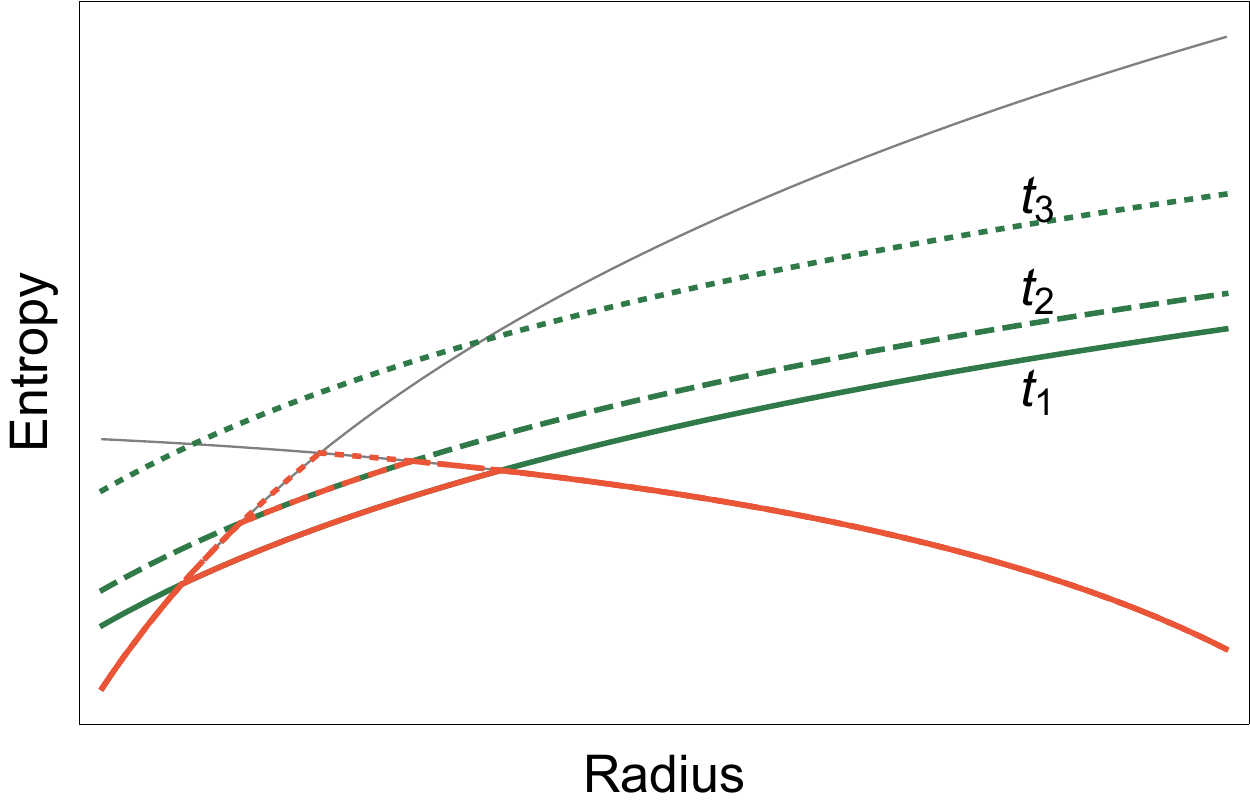}
}
\caption{\small{\textit{(a) The entropy of a subsystem as a function of time (red solid line). The green dashed line is the temperature dependent term, which scales as $\cosh{2 \pi t /\beta}$ while the horizontal lines correspond to the time independent terms in the formula for the entropy, Eq.~\eqref{eq:entropy-bh-gravity}. 
(b) The entropy as a function of the size of the subsystem, $r$. The gray lines correspond to the time independent terms in Eq.~\eqref{eq:entropy-bh-gravity}, and are monotonically increasing/decreasing with $r$. The green lines correspond to the time dependent part, which increases as time increases ($t_1 < t_2 < t_3$). At smaller times, this contributes non-trivially to the entropy (green solid and green dashed lines), but is irrelevant at later times (green dotted line).
}}}
\label{pic:entropy}
\end{centering}
\end{figure}

It is straightforward to see that for a generic $r$ and boundary entropies $g_{L}$ and $g_{R}$ the entanglement entropy, as a function of time, has a rising part (coming from the $\cosh(2\pi t/\beta)$ and a constant part (coming from the other two candidates, which ever is lower, for a given $r$). This is consistent with the spirit of a Page curve (shown in Fig.~\ref{pic:entropy-time}) and an underlying unitarity of the dynamics, which prohibits entropy from monotonically increasing with time. Another interesting feature is the behavior of entropy as a function of the radius $r$. Since the system has a finite length $L$, we expect the entropy to start and end at zero as we vary the length of the subsystem (in one limit, the subsystem is the full system and in other, the complement is the full system). However for given $r_I, r_O$, the time dependent $\cosh(2\pi t/\beta)$ part can contribute, at least for small $t$. This makes the entropy as a function of $r$ also depend on time, at least for initial times. The situation for three different times $t_1 < t_2 < t_3$ is shown in Fig.~\ref{pic:entropy-radius}. However, it deserves to be noticed that if $r$ is close enough to $r_{O}$ or $r_{I}$ or if the boundary entropies $g_{L}$ and $g_{R}$ are small enough, the Hartman-Maldacena surface is subdominant even at the very beginning. Hence in these cases we will see a constant Page curve. This is consistent with the observation in \cite{Geng:2020fxl} that to see the time dependent part of the Page curve, the subsystem Hilbert space has to be large enough.

\section{Conclusions and future directions}\label{sec:conclusion}

In this paper, we constructed a conformal field theory model for information transfer between two braneworld black holes. The black holes are embedded in two Karch-Randall branes, induced from a bulk AdS$_{3}$ planar black hole geometry. The field theory
is a two-dimensional CFT living on a strip, with generically different conformal boundary conditions on the two edges. Using the AdS/CFT correspondence, these edges are dualized to two Karch-Randall branes.

We computed an entanglement entropy similar to the L/R entanglement entropy introduced in \cite{Geng:2020fxl} to characterize the communication between the two black holes. The calculations were performed on both the field theory side and the gravitational side. We found perfect agreement between these descriptions, 
and a Page curve in accord with 
the unitarity of the underlying dynamics was obtained.

The setup introduced in \cite{Geng:2020fxl} can be obtained from the brane configuration discussed here by sending the spatial displacement $L$ of the brane endpoints on the conformal boundary to zero.
It is hence a natural question whether one can take the $L\rightarrow 0$ limit of our results in a meaningful way.
However, as far as our results for the entanglement entropy are concerned, this limit is obstructed as it requires sending also the regulator $\epsilon$ to zero.
Moreover, we notice that in this limit the boundary conformal field theory degenerates to a one-dimensional theory living on the brane-endpoint.
It is well known that such a theory, namely conformal quantum mechanics, generically does not have normalizable energy eigenstates \cite{deAlfaro:1976vlx,BrittoPacumio:1999ax}.
We therefore do not expect that the $L=0$ case can be straightforwardly recovered by na\"ivly taking the formal $L\rightarrow 0$ limit of our expressions.

An interesting story appears in the study of the zero temperature case. To perform this calculation on the field theory side, we used a conformal transformation to eliminate the local scaling behaviour of the boundary condition changing operator. This reduced the question 
to a one-point function of a primary operator on the twisted upper half plane. 
Then, as it is standard in 
holographic CFTs, 
we invoked vacuum dominance and
simply maximized over BOE channels for the two boundaries.
On the gravity side, we followed the double-holographically realized quantum extremal surface prescription to compute this entanglement entropy in the AdS$_{3}$ geometry with two Karch-Randall branes. We found a precise match with the field theory result. This provides an extension of the AdS$_{3}$/BCFT$_{2}$ correspondence.\footnote{It deserves to be emphasized that for the KR branes to model conformal boundary conditions it is essential that the branes are non-dynamical (no DGP term \cite{Dvali:2000hr}) and that there are no extra degrees of freedom localized on the branes. Otherwise the branes can absorb and store energy from the bulk violating the conformal boundary conditions. We thank the referee for pointing this out.} 
However, we emphasize that it would be interesting to repeat the analyses of \cite{Hartman:2013mia, Sully:2020pza} and determine the microscopic conditions which ensure the vacuum dominance on the twisted upper half plane.
Presumably, this would involve constraints on the spectrum of excitations induced by the presence of distinct boundaries.

We also discovered that in the 
conical defect geometry at zero temperature, corresponding to $\Delta_\text{bcc} < c/12$, geometric constraints translate into nontrivial spectral constraints. The requirement that the brane does not self-intersect results in a gap in the allowed spectrum of bcc operators, $(\tfrac{c}{16}, \tfrac{c}{12})$.\footnote{This is consistent with the bound $h_\text{bcc} \geq c/24$ proposed in \cite{Miyaji:2021ktr}. It would be interesting to better understand the connection between the two approaches.}
Similarly, the requirement that the branes subtend the correct angle forces their tension to be negative to ensure that they do not wrap around the conical defect.
These implications from the simple holographic dual are somewhat surprising, and deserve further exploration from the microscopic viewpoint.
Along similar lines, for a generic CFT living on the twisted-upper-half-plane the matching between the two boundary channels provides an interesting bootstrap constraint.\footnote{During the final stages of our work, reference \cite{Antunes:2021qpy} appeared, which studies some aspects of the bootstrap constraint we propose here.} We leave these problems to future work. 
    
Moreover, as a consequence of the mismatching boundary condition or equivalently of the different brane-tensions, the two branes intersect at an angle in the bulk, as illustrated in Fig.~\ref{pic:defectbrane}.
Typically, in theories which allow for a quantum gravity UV-completion, there can be additional light degrees of freedom localized at the intersection of branes.
For example, in string theory intersecting D-branes give rise to charged matter states localized at the brane intersection, a fact which was often used in phenomenological model building (see e.g.~\cite{Ibanez:2012zz}).
Such states are evidently not visible in our low-energy effective action and therefore not included in our computation, but one might speculate whether their inclusion could have a qualitative effect on our results.
    
Finally, it should be emphasized once more that our extension of the AdS$_{3}$/BCFT$_{2}$ correspondence has another interesting feature. The energy sourced by a boundary condition changing operator warps the bulk in a scaling-dimension dependent way, rather than simply cutting out portions of vacuum AdS$_{3}$ with Karch-Randall branes. This may therefore be a profitable setting in which we can further study the connection between information transfer, gravitational backreaction, and the black hole interior.
\pagebreak
\section*{Acknowledgements}
We would like to thank Andreas Karch, Mark Van Raamsdonk and Lisa Randall for comments on the draft, and Hong Liu, Suvrat Raju, Jamie Sully, Raman Sundrum, Tadashi Takayanagi and Chirstoph Uhlemann for helpful discussions. SL is supported by the US National Science Foundation grant NSF PHY-1915071. RKM is supported by the National Science Foundation under Grant No. NSF PHY-1748958 and NSF PHY-1915071. DW is supported by an International Doctoral Fellowship from the University of British Columbia. HG is very grateful to his parents and recommenders.

\bibliographystyle{JHEP}
\bibliography{BCFTtwoBH}
\end{document}